\documentclass[prl,twocolumn,showpacs,amsmath,amssymb]{revtex4}
\usepackage{amsfonts}
\usepackage{graphicx}
\usepackage{amsmath}
\usepackage{times}
\usepackage{amssymb}
\usepackage{color}
\usepackage[colorlinks,bookmarks=false,citecolor=blue,linkcolor=red,urlcolor=blue]{hyperref}
\usepackage{changes}

\begin{document}

\title{Anderson tower of states and nematic order of spin-1 bosonic atoms on a 2D lattice}

\author{Laurent de Forges de Parny$^{1,2}$, Hongyu Yang$^1$, and Fr\'ed\'eric Mila$^1$}
\affiliation{$^1$ Institut de th\'eorie des ph\'enom\`enes physiques, \'Ecole Polytechnique F\'ed\'erale de Lausanne (EPFL), CH-1015 Lausanne, Switzerland}
\affiliation{$^2$ Laboratoire de Physique, \'Ecole Normale Sup\'erieure de Lyon, 46 All\'ee d'Italie, 69364 Lyon Cedex 07, France}

\date{\today}

\begin{abstract}
We investigate the structure of the spectrum of antiferromagnetically coupled spin-1 bosons on a square lattice using
degenerate perturbation theory and exact diagonalizations of finite clusters. We show that
the superfluid phase develops an Anderson tower of states typical of nematic long-range order with broken $SU(2)$ 
symmetry.
We further show that this order persists into the Mott insulating phase down to zero hopping for one boson per site, 
and down to a  critical hopping for two bosons per site, in agreement with mean-field and Quantum Monte Carlo results. 
The connection with the transition between a fragmented condensate and a polar one in a single trap is briefly discussed.
\end{abstract}


\maketitle

\section{Introduction}
Spinor Bose gases have been the subject of a very intensive activity over the past fifteen years, both experimentally
and theoretically \cite{stamper-kurn2001,kawaguchi2012,lewenstein2007,stamper-kurn2013}. 
For spin-1 bosons, the spin-spin interaction can be ferromagnetic or 
antiferromagnetic depending on the relative scattering lengths in the $S=0$ and $S=2$ channels, leading in a harmonic trap 
to a ferromagnetic or to a singlet condensate\cite{ho1998,ohmi1998}. When an optical lattice is introduced, the system can in addition turn into a Mott insulator
at commensurate filling if the tunneling amplitude is small enough as compared to the on-site repulsion. In the single-band approximation 
at each site, such systems can be described by the Bose-Hubbard Hamiltonian\cite{jaksch1998,imambekov2003}:
\begin{eqnarray}
{\mathcal H} &=& -t\sum_{\langle i,j\rangle,\sigma} (a^\dagger_{i,\sigma} a^{\vphantom{\dagger}}_{j,\sigma} +\text {H.c.})
+\frac{U_0}{2}\sum_i n_i (n_i-1) \nonumber \\
&+&\frac{U_2}{2} \sum_i (\vec S_i^2-2n_i)
\label{Hamiltonian}
\end{eqnarray}
where $\langle i,j\rangle$ stands for pairs of nearest neighbors, $\sigma=-1,0,1$ is the spin, $a^\dagger_{i,\sigma}$ and
$a^{\vphantom{\dagger}}_{i,\sigma}$ are creation and annihilation operators of spin-1 bosons at site $i$, while 
${n}_i = \sum_{\sigma} {n}_{ \sigma i} = \sum_{\sigma} a^\dagger_{\sigma i} a^{\phantom{\dagger}}_{\sigma i}$  
and  $\vec S_i$  are the density and spin operators at site $i$. The parameters of this model are the tunneling amplitude $t>0$, the on-site repulsion
$U_0>0$,  and the on-site spin-spin interaction $U_2$, which is positive (negative) for antiferromagnetic (ferromagnetic) interactions.

The mean-field phase diagram of the antiferromagnetic version of the model has been mapped out quite some 
time ago by A. Imambekov \textit{et al.} \cite{imambekov2003}, who
found that the odd-density Mott insulating phases are completely nematic while the even-density ones undergo a 
transition from a non-magnetic
singlet phase to a nematic phase upon increasing the ratio $t/U_0$. In view of the competing orders (such as valence-bond solid
order reported in 1D\cite{zhou2003,rizzi2005,apaja2006}), this result clearly calls for further investigations beyond mean-field. The first attempt has been done recently using Quantum Monte Carlo, which has no minus sign problem for this type of bosonic Hamiltonian \cite{deforges2013}. This investigation 
revealed the presence of a local quadrupolar moment in the entire Mott insulating phase with one boson per site, while
a  local quadrupolar moment only develops for large enough hopping in the Mott insulating phase with two bosons per site.
This is consistent with the mean-field phase diagram, but one should keep in mind that the numerical demonstration of 
nematic long-range order would require an investigation of quadrupolar correlations, which was beyond the scope of Ref.~\cite{deforges2013}.
So further work is definitely needed to check the presence of nematic 
long-range order in the phase diagram of the model of Eq.\ref{Hamiltonian}. 

In this Letter, we show that the superfluid phase of spin-1 bosons with antiferromagnetic interactions indeed 
develops true nematic long-range order in the presence of a lattice.
This conclusion is based on a careful investigation of the excitation spectrum of the model using degenerate perturbation theory in the limit $U_0=0,U_2/t\rightarrow 0$
and exact diagonalizations of finite clusters away from that limit. 
The key observation is that, in the presence of a lattice, the spectrum acquires the structure of an  
Anderson tower of states, \textit{i.e.} a family of low-lying states whose energy collapses onto that of the ground state
in the thermodynamic limit, and that all these states have even values of the total spin, so that
polar states (and not antiferromagnetic states) can be reconstructed as linear combinations of 
degenerate ground states. Exact diagonalizations are further used to show that this structure persists in the Mott insulating 
phase as long as nematic order is present, leading to an alternative determination of the singlet-nematic 
transition in the $S=2$ Mott insulator. 

Let us start by solving the problem analytically in the limit $U_0=0,U_2/t\rightarrow 0$ which,
as we shall show later, turns out to be representative of the general case.  Let us denote by $N_s$ the number of sites
 and by $N$ the number of bosons. In the non-interacting case ($U_0=U_2=0$), the bosons condense 
 in the $\vec k=\vec 0$ state, but since there is no magnetic interaction,
the spin is irrelevant, and the ground state is vastly degenerate. The ground states are given by
\begin{equation}
\vert \psi_{n_{-1},n_0,n_1} \rangle = \prod_{\sigma} \frac{a^{\dagger n_\sigma}_{\vec k=\vec 0,\sigma}}{\sqrt{n_\sigma!}} \vert 0 \rangle
\label{degenerate_GS}
\end{equation}
with $\sum_\sigma n_\sigma=N$ and $a^\dagger_{\vec k=\vec 0,\sigma}=(1/\sqrt{N_s}) \sum_{i=1}^{N_s} a^\dagger_{i,\sigma}$.
The degeneracy is equal to $(N+1)(N+2)/2$. 

Let us now consider the effect of $U_2$. If $U_2/t$ is small, we can use degenerate perturbation theory, which means
that we must diagonalize $\sum_i \vec S_i^2$ in the subspace spanned by the degenerate ground states of Eq.(\ref{degenerate_GS}). Now, this operator
commutes with the square of the total spin $\vec S_{\text{tot}}=\sum_i \vec S_i$. So, in the basis of the eigenstates of 
$\vec S_{\text{tot}}^2$, the matrix of $\sum_i \vec S_i^2$ is diagonal, and the problem reduces to the evaluation of the
expectation value of $\sum_i \vec S_i^2$ in the eigenstates of $\vec S_{\text{tot}}^2$. Since $\sum_i \vec S_i^2$
also commutes with the components of $\vec S_{\text{tot}}$, hence with $S_{\text{tot}}^-$ and $S_{\text{tot}}^+$, the expectation value
in a state $\vert S_{\text{tot}},m\rangle$ does not depend on $m$, and it is sufficient to calculate it in one member of the
family, for instance $\vert S_{\text{tot}},m=S_{\text{tot}}\rangle$. The calculation of the expectation value of $\vec S_i^2$
in this state can be done analytically (see Supplemental Material\cite{SM}), leading to:
\begin{equation}
\langle \vec S_i^2 \rangle_{S_{\text{tot}}} = \frac{2N(N_s-1)}{N_s^2}+\frac{1}{N_s^2}S_{\text{tot}}(S_{\text{tot}}+1)
\label{expectation_Si2}
\end{equation}
As anticipated, $\langle \vec S_i^2 \rangle$ is only a function of $S_{\text{tot}}$. This dependence turns out to take the very
simple form $S_{\text{tot}}(S_{\text{tot}}+1)$, but this is by no means a trivial result in the sense that $\sum_i \vec S_i^2$ is not
simply related to  $\vec S_{\text{tot}}^2$. In fact, $\vec S_{\text{tot}}^2=\sum_i \vec S_i^2+\sum_{i\neq j}\vec S_i\cdot \vec S_j$,
and the expectation value of $\vec S_i\cdot \vec S_j$ in $\vert S_{\text{tot}},m=S_{\text{tot}}\rangle$ does not vanish but is given by
\begin{equation}
\langle \vec S_i \cdot \vec S_j\rangle_{S_{\text{tot}}} = -\frac{2N}{N_s^2}+\frac{1}{N_s^2}S_{\text{tot}}(S_{\text{tot}}+1)
\label{expectation_SiSj}
\end{equation}
Eq.(\ref{expectation_Si2}) implies in particular that, in the total singlet, and in the thermodynamic limit $N,N_s\rightarrow +\infty$, 
$\rho=N/N_s$ fixed, the local value of the square of the spin is given by
\begin{equation}
\langle \vec S_i^2 \rangle_{S_{\text{tot}}=0} = 2\rho
\label{expectation_Si2_thermodynamic limit}
\end{equation}
where $\rho$ is the boson density. Contrary to what one might naively expect, this limiting value is {\it not} of the form $S(S+1)$
for some integer $S$. It is however in good agreement with the QMC results obtained for one and two bosons per site \cite{deforges2013}.
Finally, let us emphasize that, in Eq.~(\ref{expectation_Si2}),  $S_{\text{tot}}$ can only take even values because it corresponds
to the total spin of spin-1 bosons in a single mode, the $\vec k = \vec 0$ one. 

\begin{figure}
\begin{center}
\includegraphics[width=1\columnwidth]{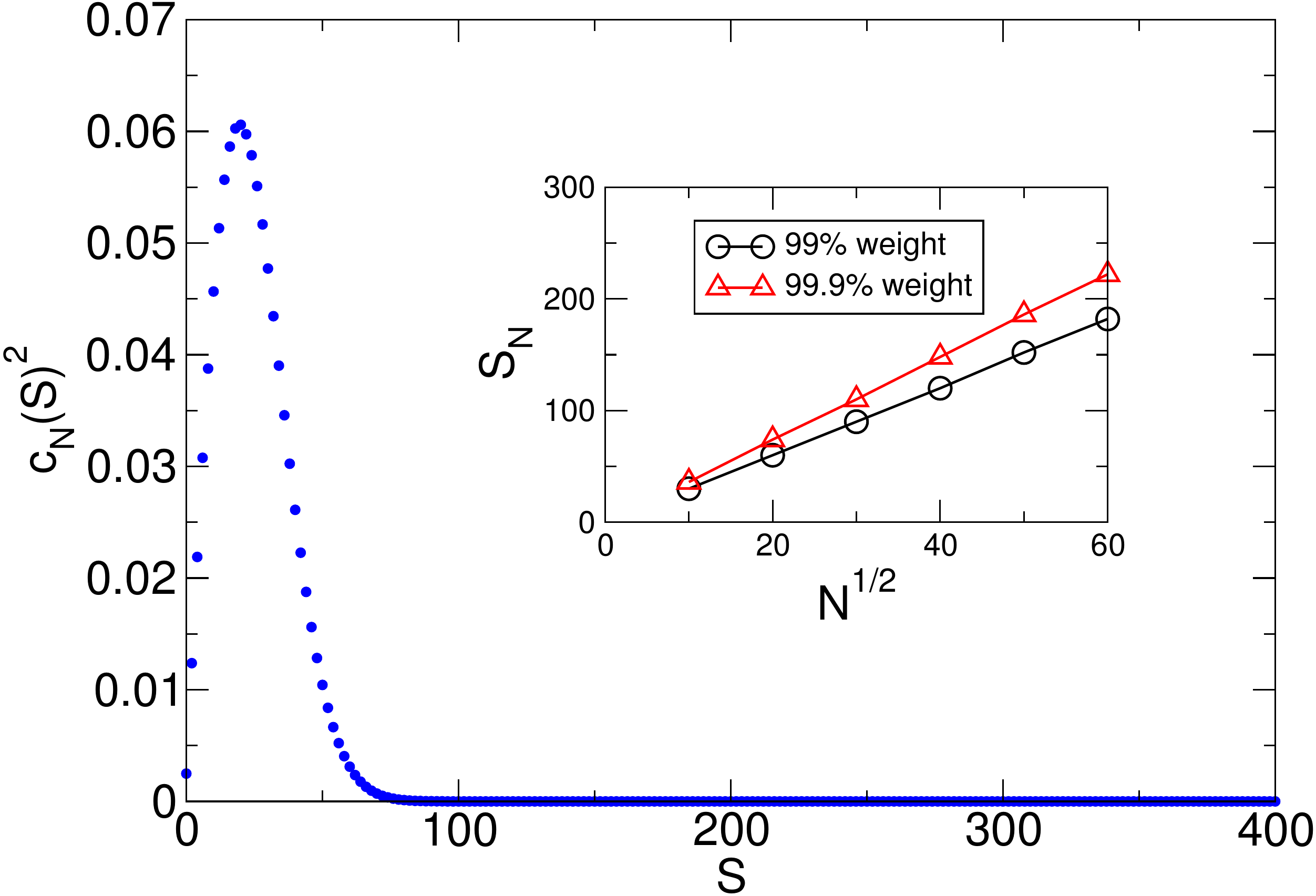}
\caption{(Color online) Squares $c_N(S)^2$ of the coefficients of the expansion of the polar state $\vert \psi_{0,N,0}\rangle$ in the eigenstates $\vert S,m=0\rangle$ of $\vec S_{\text{tot}}^2$ as a function of $S$
for $N=400$. Inset: value of the spin $S_N$ up to which one has to sum to satisfy the sum rule $\sum_S \vert c_N(S)\vert^2=1$ to a given 
accuracy. It scales as $\sqrt{N}$.
}
\label{fig_states}
\end{center}
\end{figure}

Coming back to the Hamiltonian of Eq.(\ref{Hamiltonian}) in the limit $U_0=0,U_2/t\rightarrow 0$, the low-energy spectrum 
is thus given by
\begin{equation}
E_{S_{\text{tot}}} = -4tN-\frac{N}{N_s} U_2+\frac{U_2}{2}\frac{1}{N_s}S_{\text{tot}}(S_{\text{tot}}+1)
\label{spectrum_tower_of_states}
\end{equation}
where the first term is the energy of the non-interacting condensate.
The important property is that the slope is proportional to $1/N_s$ and tends to zero in the thermodynamic limit,
leading to a quasi-degenerate ground state. In quantum antiferromagnets, this property goes under the name of Anderson's tower of
states\cite{anderson1952,bernu1992,bernu1994,lhuillier2005,misguich2007}: on the basis of the low-lying states of this tower, 
it is possible to reconstruct a wave function very close to the N\'eel state with
spins up on one sublattice and down on the other sublattice whose energy is very low and scales to the ground
state energy when the system size increases, so that the appearance of a tower of states in the low-energy spectrum
indicates that the SU(2) symmetry is spontaneously broken in the ground state in favor of antiferromagnetism. 

Note that the tower of states remains a well defined concept as long as the energies of the states building this tower
are well separated from those of the elementary excitations, which also tend to the ground state energy in ordered systems.
In the present calculation, which is performed in the $U_2/t \rightarrow 0$ limit, this is clearly true since the elementary 
excitations consist in exciting a particle out of the condensate with an energy of order $t/N_s$ for small wave vector, while 
the states of the tower have energies of order $U_2/N_s$.

Now, using only states with small values of $S_{\text{tot}}$, it is possible to reconstruct almost exactly
polar states. Indeed, all polar states are related by a rotation to $\vert \psi_{0,N,0} \rangle \propto a^{\dagger N}_{\vec k=\vec 0,0}\vert 0 \rangle$, and this state can be expanded in the basis of eigenstates of $\vec S_{\text{tot}}^2$ as
\begin{equation*}
\vert \psi_{0,N,0} \rangle = \sum_{S=0,2,...,N} c_N(S) \vert S,m=0\rangle
\label{decomposition_nematic_state}
\end{equation*}
The coefficients $c_N(S)$ can be determined analytically\cite{SM} and are given by
\begin{equation}
c_N(S)=\sqrt{\frac{(2S+1)\,N!}{(N-S)!!\,(N+S+1)!!}}
\label{coefficients}
\end{equation}
These coefficients only take significant values up to $S=O(\sqrt{N})$: as shown in Fig. \ref{fig_states}, the maximum value of the spin
$S_N$ up to which one has to sum to satisfy the sum rule $\sum_S \vert c_N(S)\vert^2=1$ to a given 
accuracy scales as $\sqrt{N}$. As a consequence, the polar state $\vert \psi_{0,N,0}\rangle$ has an energy 
per site that scales to the ground state one as $1/N_s$ in the thermodynamic limit. It is thus
a ground state in that limit, which proves the presence of long-range nematic order.
\begin{figure}
\begin{center}
\includegraphics[height=6 cm]{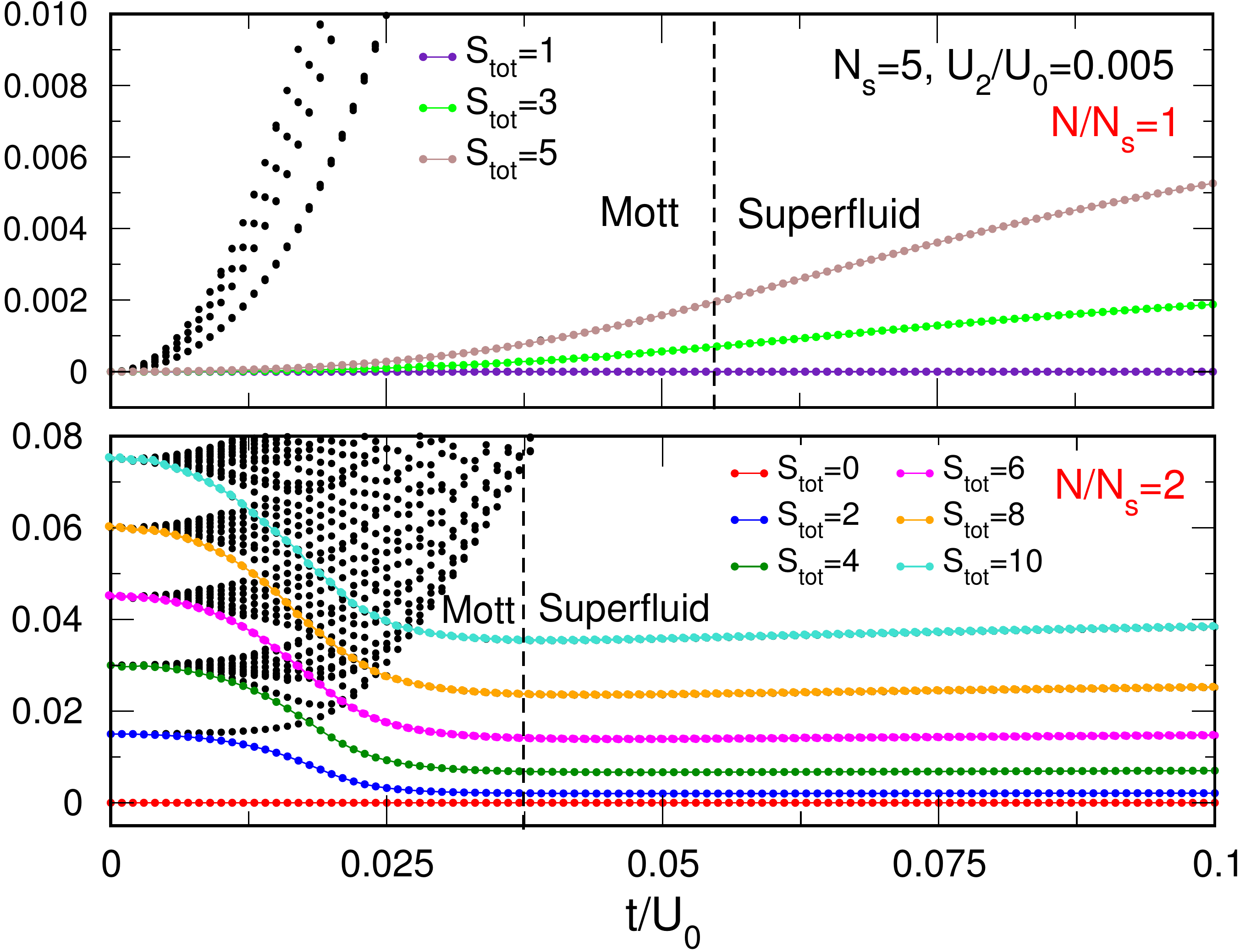}
\caption{(Color online) Low energy spectra for 5 sites with $N/N_s=1$ (up) and $N/N_s=2$ (bottom) with $U_2/U_0=0.005$. The energies
are in unit  of $U_0$ and measured from the ground state.
The vertical dashed line indicates the Mott-superfluid transition according to the QMC simulations\cite{deforges2013}.
}
\label{fig_spectrum_5sites}
\end{center}
\end{figure}
These results establish that, in the limit $U_0=0,U_2/t\rightarrow 0$, the SU(2) symmetry of the superfluid 
ground state of spin-1 bosons on a lattice is spontaneously broken in favor of nematic order. 

It is instructive to compare these results to the case of spin-1 bosons in a trap\cite{ho2000,koashi2000,mueller2006,Tasaki,gerbier2013}. 
In the single-mode approximation, the Hamiltonian reads
\begin{equation}
{\cal H}=
\frac{U_s}{2N} \vec S^2
\end{equation}
where $U_s$ is the spin interaction energy per atom. 
The ground state is a non-degenerate singlet with energy $E_0=0$, and the excitation energies are given
by $E_S=U_s/(2N)S(S+1)$,  $S=2,4,...$. The structure of this spectrum is similar to that of Eq.(\ref{spectrum_tower_of_states}),
with in particular a slope that goes to zero as $1/N$ in the thermodynamic limit. Accordingly, the consequences 
in that limit are very similar: As discussed in Refs.\cite{ho2000,koashi2000,mueller2006,Tasaki,gerbier2013}, spontaneous symmetry 
breaking takes place since a polar state can be stabilized for infinitesimal quadratic  Zeeman coupling coupling in the 
thermodynamic limit. The decomposition of the polar state into the angular momentum basis also takes a very similar form,
our analytical result of Eq.(\ref{coefficients}) corresponding to the large-$q$ limit of the result of Ref.\cite{gerbier2013}, where the 
transition between a fragmented condensate and a polar state induced by a quadratic  Zeeman coupling $q$ has been investigated in detail.

Let us now turn to
the general case. The phase diagram has been previously studied using mean-field theory \cite{imambekov2003,kimura2005,pai2008}, variational Monte Carlo method \cite{toga2012} and quantum Monte Carlo simulations \cite{batrouni2009, deforges2013}, and
some exact results have been established\cite{katsura2013}.
These methods have led to the conclusion that, for integer filling, there is a superfluid-insulator Mott transition upon increasing 
$U_0/t$, and the insulating state is always nematic for odd filling while there is an additional nematic-singlet transition upon
further increasing $U_0/t$ for even filling. Now that we have analytically demonstrated that the superfluid state is nematic in the limit $U_0/t=0$
on the basis of the structure of the low-energy spectrum, it is natural to ask to which extent this structure persists away from that
limit. 
For that purpose, we have performed exact diagonalizations of finite-size clusters for one and two bosons per site, with up to 
10 and 8 sites respectively. The low energy spectra for 5 sites are depicted in Fig. \ref{fig_spectrum_5sites} as a function of $t/U_0$. 
In both cases, the total spin $S_{\text{tot}}$ changes by 2 from one state to the next, and the energy of a state measured from the ground state is proportional to $S_{\text{tot}}(S_{\text{tot}}+1)$, with a coefficient that tends to $U_2/2N_s$ 
in the large $t/U_0$ limit, in agreement with Eq.(\ref{spectrum_tower_of_states}).
This structure persists below the superfluid-insulator transition without any hint that the system undergoes a phase transition. This suggests that the nematic order predicted previously in 
the insulating phases is continuously related to the nematic order we have established in 
the $U_0=0,U_2/t\rightarrow 0$ limit. Upon further reducing the ratio $t/U_0$, the structure of the tower of states remains essentially unaffected for one boson per site, 
but a series of level crossings leads to a completely different spectrum for two bosons per site which signals a nematic-singlet transtion. The same structure has been observed on larger clusters\cite{SM}.

\begin{figure}
\begin{center}
\includegraphics[width=1\columnwidth]{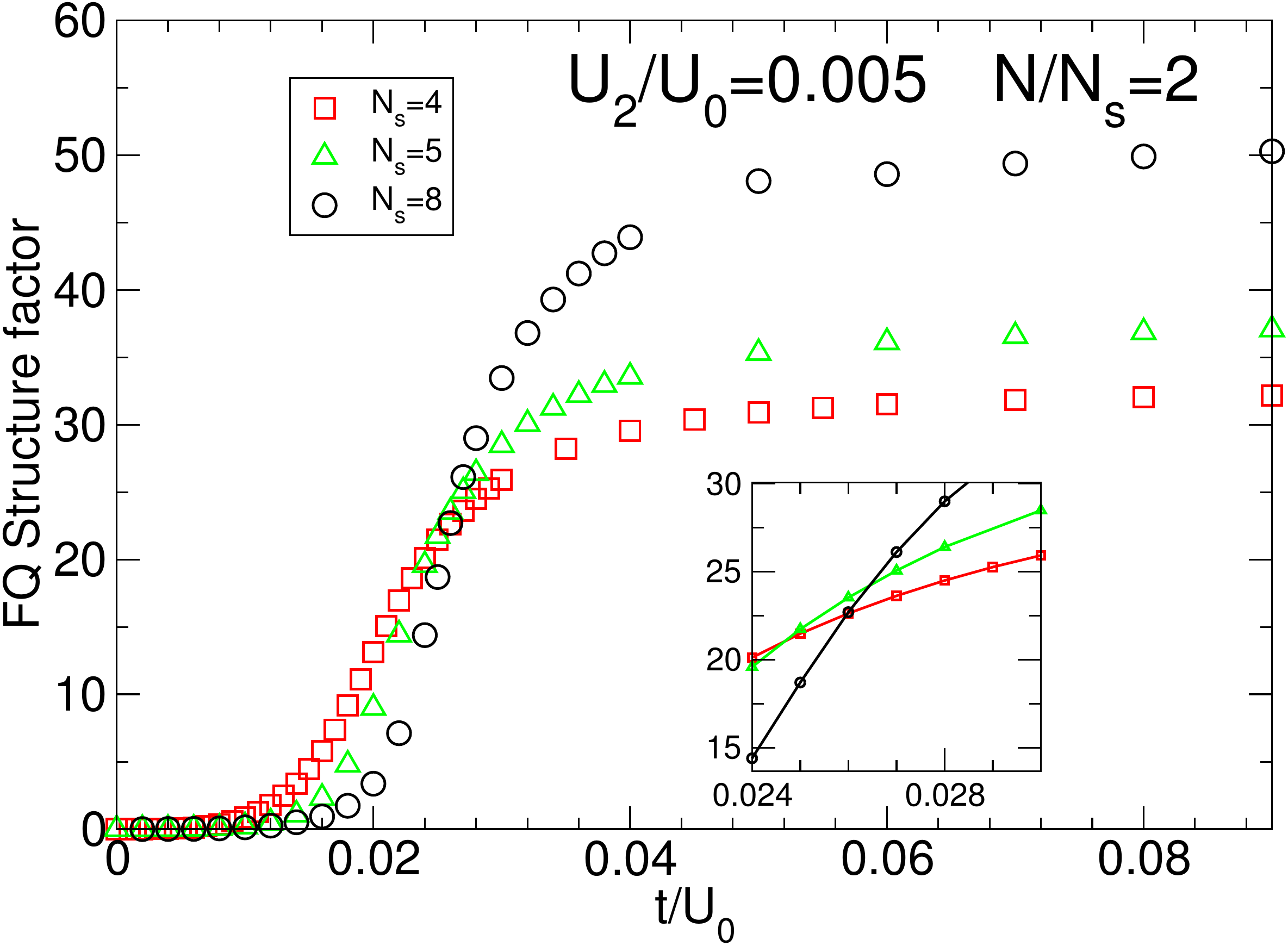}
\caption{(Color online) Ferromagnetic quadrupolar structure factor for 4, 5 and 8-site clusters with two bosons per site.
}
\label{fig_structure_factor}
\end{center}
\end{figure}

In the Mott-insulating phase, the identification of the order as nematic, and {\it not} antiferromagnetic, actually deserves special attention
since, for antiferromagnetic coupled spins, one might in general expect simple N\'eel order on a bipartite lattice.  In exact diagonalizations, antiferromagnetic and nematic order can be distinguished by the quantum numbers that
appear in the tower of states. For N\'eel order, all values of $S_{\text{tot}}$ are represented in the tower of states because,
to reconstruct the N\'eel state with up spins on one sublattice and down spins on the other one, one needs states with both even and 
odd total spin whereas, for quadrupolar order, one only needs states with even total spin\cite{shannon2006,penc2011}. In view of the analytical results of the $U_0=0,U_2/t\rightarrow 0$ limit, we expect by continuity the low-lying states calculated by exact diagonalizations away from that limit to carry only total spin. We have explicitly checked this to be the case for the 5 site cluster. So the fact that only even 
steps in $S_{\text{tot}}$ appear in the tower of states is an additional confirmation that, in the regions of the 
Mott insulating phases with spontaneously broken SU(2) symmetry 
the order is indeed nematic.

As an independent confirmation, we have calculated the ferroquadrupolar structure factor $S^Q(\vec k = \vec 0)$, with  $S^Q(\vec k) =\sum_j  \exp(i \vec k \cdot \vec r_j ) \left< \vec Q_0 \cdot \vec Q_j \right> $, where the quadrupolar operator is 
 defined by $\vec Q=[(S^x)^2-(S^y)^2,1/\sqrt{3}(2(S^z)^2-(S^x)^2-(S^y)^2),S^xS^y + S^yS^x,S^yS^z + S^zS^y,S^zS^x + S^xS^z]$.
For two bosons per site, as can be seen in Fig. \ref{fig_structure_factor}, it increases with the size for 
large enough $t/U_0$, and it decreases with the size for small enough $t/U_0$. The crossing point can be taken
as an approximation of the transition to nematic order\cite{SM}, and the critical 
value $t_c/U_0\sim0.026$
is in excellent agreement with the QMC estimate based on
the development of a local quadrupolar moment. 

In principle, it is possible to locate the singlet-nematic transition just by investigating the spin gap, 
which is expected to  be finite in the singlet phase and to scale to zero in the nematic phase. It turns out that this is 
not very accurate   for the sizes accessible with exact diagonalizations, and only a very rough estimate of the transition
can be obtained along these lines\cite{SM}. This estimation however is still consistent with other estimates.

\begin{figure}
\begin{center}
\includegraphics[height=5 cm]{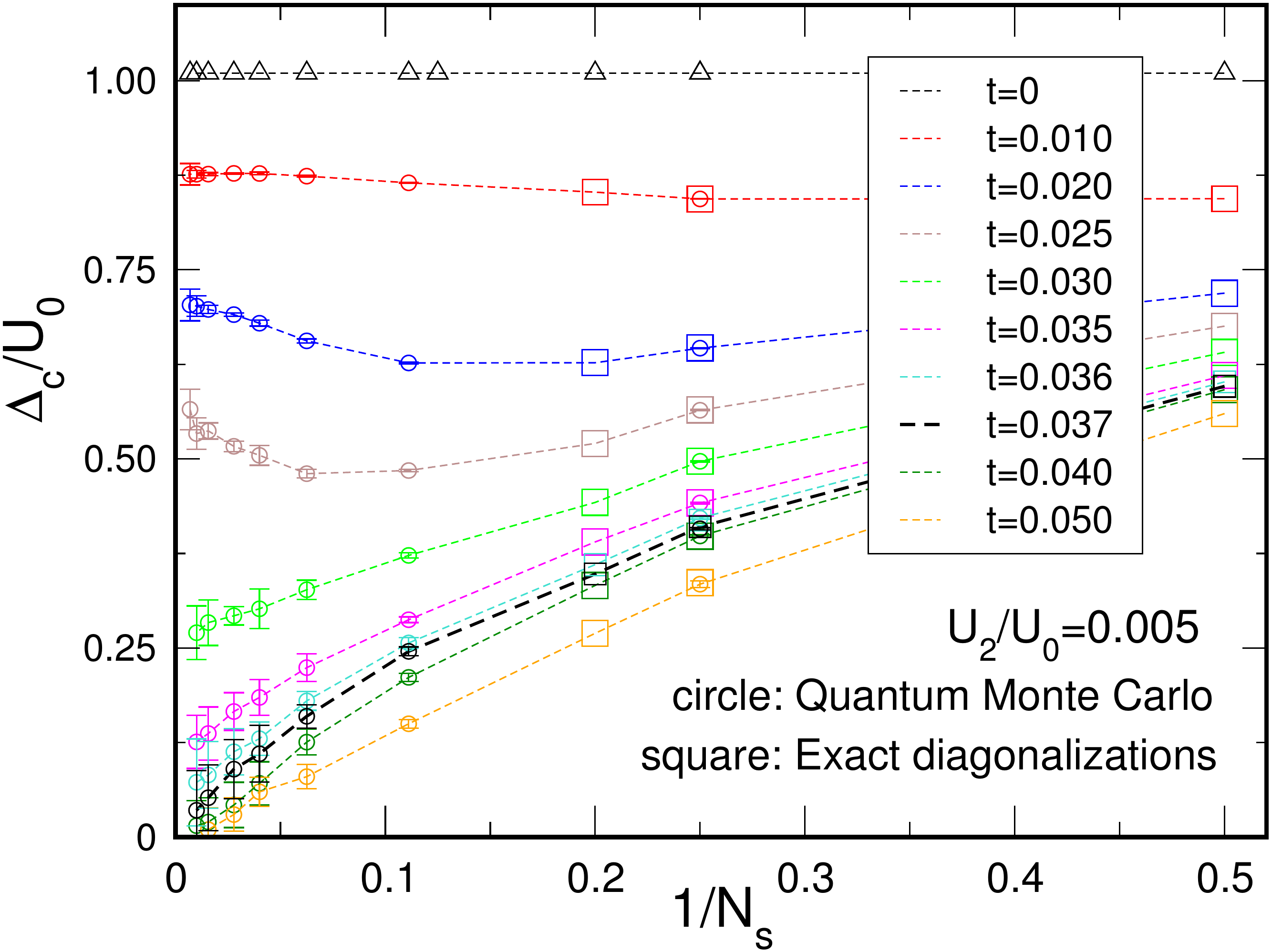}
\caption{(Color online) Charge gap as a function of the inverse number of sites for various values of $t/U_0$. It opens at $t_c /U_0 \simeq 0.037$, which marks the superfluid-insulator transition.
}
\label{fig_GAPextrapol_Ut0p005}
\end{center}
\end{figure}

Finally, we have attempted to locate the superfluid-insulator
transition, which, as usual, corresponds to the opening of the charge gap defined by
$
\Delta_c = E(N+1)+E(N-1)-2E(N)
$
As for the singlet-triplet gap, the results are consistent with the QMC estimate ($t_c / U_0 \simeq 0.037$), but the sizes accessible to exact diagonalizations
do not lead to a very precise estimate. It is only thanks to QMC calculations, with SGF algorithm \cite{rousseau2008}, of the energies for larger system sizes that this
criterion can be shown to coincide with the appearance of a superfluid stiffness\cite{deforges2013} (see Fig. \ref{fig_GAPextrapol_Ut0p005}) . Note that, for 
$U_2/U_0 =0.005$, the superfluid-Mott insulator transition is well separated from the singlet-nematic transition. Increasing $U_2$ pushes the singlet-nematic transition to larger critical value, and when $U_2/U_0 \sim 0.1 $, there is only one transition from the singlet Mott phase to the nematic superfluid left\cite{SM}.

To summarize, let us put the present results in perspective. Except in one dimension, where Density Matrix Renormalization 
Group can be used\cite{rizzi2005,bergkvist2006,rodriguez2011}, 
the investigation of lattice bosonic models is largely dominated
by QMC, and rightly so since, due to the absence of minus sign problem in many cases, extremely accurate results can be 
obtained on very large system sizes. Yet, as demonstrated in the present Letter, investigating the excitation spectrum of the
model with analytical tools if possible, or with exact diagonalizations of small clusters, can lead to very interesting insight into
the properties of the system, even if the sizes accessible are much smaller than with QMC. In the present case, the structure
of the low-energy spectrum, which consists of an Anderson tower of state in a large portion of the phase diagram, is an 
extremely fruitful piece of information. In particular, it has led to the demonstration that, in the superfluid phase of 
spin-1 bosons on a lattice, the SU(2) symmetry is spontaneously broken, by contrast to the case of bosons in a single mode, 
which require an SU(2) symmetry-breaking interaction to build a polar condensate. It will be very interesting to investigate
the implications of this result on the dynamics of spinor condensates\cite{lamacraft2010} in the presence of a lattice.

\begin{acknowledgements} 
We thank G. Batrouni, F. H\'ebert, A. L\"auchli and V. G. Rousseau for useful discussions, and T. Roscilde for his critical reading
of the manuscript. We are especially indebted to F. Gerbier for insightful remarks on the thermodynamic limit in a single trap, and
to an anonymous referee for a simpler proof of Eq.\ref{expectation_Si2}. This work has been supported by
the Swiss National Fund.
\end{acknowledgements}

\bibliographystyle{apsrev4-1}
\bibliography{refs,comments}

\begin{thebibliography}{34}%
\makeatletter
\providecommand \@ifxundefined [1]{%
 \@ifx{#1\undefined}
}%
\providecommand \@ifnum [1]{%
 \ifnum #1\expandafter \@firstoftwo
 \else \expandafter \@secondoftwo
 \fi
}%
\providecommand \@ifx [1]{%
 \ifx #1\expandafter \@firstoftwo
 \else \expandafter \@secondoftwo
 \fi
}%
\providecommand \natexlab [1]{#1}%
\providecommand \enquote  [1]{``#1''}%
\providecommand \bibnamefont  [1]{#1}%
\providecommand \bibfnamefont [1]{#1}%
\providecommand \citenamefont [1]{#1}%
\providecommand \href@noop [0]{\@secondoftwo}%
\providecommand \href [0]{\begingroup \@sanitize@url \@href}%
\providecommand \@href[1]{\@@startlink{#1}\@@href}%
\providecommand \@@href[1]{\endgroup#1\@@endlink}%
\providecommand \@sanitize@url [0]{\catcode `\\12\catcode `\$12\catcode
  `\&12\catcode `\#12\catcode `\^12\catcode `\_12\catcode `\%12\relax}%
\providecommand \@@startlink[1]{}%
\providecommand \@@endlink[0]{}%
\providecommand \url  [0]{\begingroup\@sanitize@url \@url }%
\providecommand \@url [1]{\endgroup\@href {#1}{\urlprefix }}%
\providecommand \urlprefix  [0]{URL }%
\providecommand \Eprint [0]{\href }%
\providecommand \doibase [0]{http://dx.doi.org/}%
\providecommand \selectlanguage [0]{\@gobble}%
\providecommand \bibinfo  [0]{\@secondoftwo}%
\providecommand \bibfield  [0]{\@secondoftwo}%
\providecommand \translation [1]{[#1]}%
\providecommand \BibitemOpen [0]{}%
\providecommand \bibitemStop [0]{}%
\providecommand \bibitemNoStop [0]{.\EOS\space}%
\providecommand \EOS [0]{\spacefactor3000\relax}%
\providecommand \BibitemShut  [1]{\csname bibitem#1\endcsname}%
\let\auto@bib@innerbib\@empty
\bibitem [{\citenamefont {Stamper-Kurn}\ and\ \citenamefont
  {Ketterle}(2001)}]{stamper-kurn2001}%
  \BibitemOpen
  \bibfield  {author} {\bibinfo {author} {\bibfnamefont {D.~M.}\ \bibnamefont
  {Stamper-Kurn}}\ and\ \bibinfo {author} {\bibfnamefont {W.}~\bibnamefont
  {Ketterle}},\ }\href@noop {} {\bibfield  {journal} {\bibinfo  {journal} {in
  Coherent Matter Waves, edited by R. Kaiser, C. Westbrook, and F. David
  (Springer-Verlag, New York), Chap. 2, pp. 137–218}\ } (\bibinfo {year}
  {2001})}\BibitemShut {NoStop}%
\bibitem [{\citenamefont {Kawaguchi}\ and\ \citenamefont
  {Ueda}(2012)}]{kawaguchi2012}%
  \BibitemOpen
  \bibfield  {author} {\bibinfo {author} {\bibfnamefont {Y.}~\bibnamefont
  {Kawaguchi}}\ and\ \bibinfo {author} {\bibfnamefont {M.}~\bibnamefont
  {Ueda}},\ }\href {\doibase 10.1016/j.physrep.2012.07.005} {\bibfield
  {journal} {\bibinfo  {journal} {Physics Reports}\ }\textbf {\bibinfo {volume}
  {520}},\ \bibinfo {pages} {253 } (\bibinfo {year} {2012})}\BibitemShut
  {NoStop}%
\bibitem [{\citenamefont {Lewenstein}\ \emph {et~al.}(2007)\citenamefont
  {Lewenstein}, \citenamefont {Sanpera}, \citenamefont {Ahufinger},
  \citenamefont {Damski}, \citenamefont {Sen(De)},\ and\ \citenamefont
  {Sen}}]{lewenstein2007}%
  \BibitemOpen
  \bibfield  {author} {\bibinfo {author} {\bibfnamefont {M.}~\bibnamefont
  {Lewenstein}}, \bibinfo {author} {\bibfnamefont {A.}~\bibnamefont {Sanpera}},
  \bibinfo {author} {\bibfnamefont {V.}~\bibnamefont {Ahufinger}}, \bibinfo
  {author} {\bibfnamefont {B.}~\bibnamefont {Damski}}, \bibinfo {author}
  {\bibfnamefont {A.}~\bibnamefont {Sen(De)}}, \ and\ \bibinfo {author}
  {\bibfnamefont {U.}~\bibnamefont {Sen}},\ }\href {\doibase
  10.1080/00018730701223200} {\bibfield  {journal} {\bibinfo  {journal}
  {Advances in Physics}\ }\textbf {\bibinfo {volume} {56}},\ \bibinfo {pages}
  {243} (\bibinfo {year} {2007})}\BibitemShut {NoStop}%
\bibitem [{\citenamefont {Stamper-Kurn}\ and\ \citenamefont
  {Ueda}(2013)}]{stamper-kurn2013}%
  \BibitemOpen
  \bibfield  {author} {\bibinfo {author} {\bibfnamefont {D.~M.}\ \bibnamefont
  {Stamper-Kurn}}\ and\ \bibinfo {author} {\bibfnamefont {M.}~\bibnamefont
  {Ueda}},\ }\href {\doibase 10.1103/RevModPhys.85.1191} {\bibfield  {journal}
  {\bibinfo  {journal} {Rev. Mod. Phys.}\ }\textbf {\bibinfo {volume} {85}},\
  \bibinfo {pages} {1191} (\bibinfo {year} {2013})}\BibitemShut {NoStop}%
\bibitem [{\citenamefont {Ho}(1998)}]{ho1998}%
  \BibitemOpen
  \bibfield  {author} {\bibinfo {author} {\bibfnamefont {T.-L.}\ \bibnamefont
  {Ho}},\ }\href {\doibase 10.1103/PhysRevLett.81.742} {\bibfield  {journal}
  {\bibinfo  {journal} {Phys. Rev. Lett.}\ }\textbf {\bibinfo {volume} {81}},\
  \bibinfo {pages} {742} (\bibinfo {year} {1998})}\BibitemShut {NoStop}%
\bibitem [{\citenamefont {Ohmi}\ and\ \citenamefont
  {Machida}(1998)}]{ohmi1998}%
  \BibitemOpen
  \bibfield  {author} {\bibinfo {author} {\bibfnamefont {T.}~\bibnamefont
  {Ohmi}}\ and\ \bibinfo {author} {\bibfnamefont {K.}~\bibnamefont {Machida}},\
  }\href {\doibase 10.1143/JPSJ.67.1822} {\bibfield  {journal} {\bibinfo
  {journal} {Journal of the Physical Society of Japan}\ }\textbf {\bibinfo
  {volume} {67}},\ \bibinfo {pages} {1822} (\bibinfo {year}
  {1998})}\BibitemShut {NoStop}%
\bibitem [{\citenamefont {Jaksch}\ \emph {et~al.}(1998)\citenamefont {Jaksch},
  \citenamefont {Bruder}, \citenamefont {Cirac}, \citenamefont {Gardiner},\
  and\ \citenamefont {Zoller}}]{jaksch1998}%
  \BibitemOpen
  \bibfield  {author} {\bibinfo {author} {\bibfnamefont {D.}~\bibnamefont
  {Jaksch}}, \bibinfo {author} {\bibfnamefont {C.}~\bibnamefont {Bruder}},
  \bibinfo {author} {\bibfnamefont {J.~I.}\ \bibnamefont {Cirac}}, \bibinfo
  {author} {\bibfnamefont {C.~W.}\ \bibnamefont {Gardiner}}, \ and\ \bibinfo
  {author} {\bibfnamefont {P.}~\bibnamefont {Zoller}},\ }\href {\doibase
  10.1103/PhysRevLett.81.3108} {\bibfield  {journal} {\bibinfo  {journal}
  {Phys. Rev. Lett.}\ }\textbf {\bibinfo {volume} {81}},\ \bibinfo {pages}
  {3108} (\bibinfo {year} {1998})}\BibitemShut {NoStop}%
\bibitem [{\citenamefont {Imambekov}\ \emph {et~al.}(2003)\citenamefont
  {Imambekov}, \citenamefont {Lukin},\ and\ \citenamefont
  {Demler}}]{imambekov2003}%
  \BibitemOpen
  \bibfield  {author} {\bibinfo {author} {\bibfnamefont {A.}~\bibnamefont
  {Imambekov}}, \bibinfo {author} {\bibfnamefont {M.}~\bibnamefont {Lukin}}, \
  and\ \bibinfo {author} {\bibfnamefont {E.}~\bibnamefont {Demler}},\ }\href
  {\doibase 10.1103/PhysRevA.68.063602} {\bibfield  {journal} {\bibinfo
  {journal} {Phys. Rev. A}\ }\textbf {\bibinfo {volume} {68}},\ \bibinfo
  {pages} {063602} (\bibinfo {year} {2003})}\BibitemShut {NoStop}%
\bibitem [{\citenamefont {Zhou}\ and\ \citenamefont {Snoek}(2003)}]{zhou2003}%
  \BibitemOpen
  \bibfield  {author} {\bibinfo {author} {\bibfnamefont {F.}~\bibnamefont
  {Zhou}}\ and\ \bibinfo {author} {\bibfnamefont {M.}~\bibnamefont {Snoek}},\
  }\href {\doibase 10.1016/j.aop.2003.08.009} {\bibfield  {journal} {\bibinfo
  {journal} {Annals of Physics}\ }\textbf {\bibinfo {volume} {308}},\ \bibinfo
  {pages} {692 } (\bibinfo {year} {2003})}\BibitemShut {NoStop}%
\bibitem [{\citenamefont {Rizzi}\ \emph {et~al.}(2005)\citenamefont {Rizzi},
  \citenamefont {Rossini}, \citenamefont {De~Chiara}, \citenamefont
  {Montangero},\ and\ \citenamefont {Fazio}}]{rizzi2005}%
  \BibitemOpen
  \bibfield  {author} {\bibinfo {author} {\bibfnamefont {M.}~\bibnamefont
  {Rizzi}}, \bibinfo {author} {\bibfnamefont {D.}~\bibnamefont {Rossini}},
  \bibinfo {author} {\bibfnamefont {G.}~\bibnamefont {De~Chiara}}, \bibinfo
  {author} {\bibfnamefont {S.}~\bibnamefont {Montangero}}, \ and\ \bibinfo
  {author} {\bibfnamefont {R.}~\bibnamefont {Fazio}},\ }\href {\doibase
  10.1103/PhysRevLett.95.240404} {\bibfield  {journal} {\bibinfo  {journal}
  {Phys. Rev. Lett.}\ }\textbf {\bibinfo {volume} {95}},\ \bibinfo {pages}
  {240404} (\bibinfo {year} {2005})}\BibitemShut {NoStop}%
\bibitem [{\citenamefont {Apaja}\ and\ \citenamefont
  {Sylju\aa{}sen}(2006)}]{apaja2006}%
  \BibitemOpen
  \bibfield  {author} {\bibinfo {author} {\bibfnamefont {V.}~\bibnamefont
  {Apaja}}\ and\ \bibinfo {author} {\bibfnamefont {O.~F.}\ \bibnamefont
  {Sylju\aa{}sen}},\ }\href {\doibase 10.1103/PhysRevA.74.035601} {\bibfield
  {journal} {\bibinfo  {journal} {Phys. Rev. A}\ }\textbf {\bibinfo {volume}
  {74}},\ \bibinfo {pages} {035601} (\bibinfo {year} {2006})}\BibitemShut
  {NoStop}%
\bibitem [{\citenamefont {de~Forges~de Parny}\ \emph
  {et~al.}(2013)\citenamefont {de~Forges~de Parny}, \citenamefont {H\'ebert},
  \citenamefont {Rousseau},\ and\ \citenamefont {Batrouni}}]{deforges2013}%
  \BibitemOpen
  \bibfield  {author} {\bibinfo {author} {\bibfnamefont {L.}~\bibnamefont
  {de~Forges~de Parny}}, \bibinfo {author} {\bibfnamefont {F.}~\bibnamefont
  {H\'ebert}}, \bibinfo {author} {\bibfnamefont {V.~G.}\ \bibnamefont
  {Rousseau}}, \ and\ \bibinfo {author} {\bibfnamefont {G.~G.}\ \bibnamefont
  {Batrouni}},\ }\href {\doibase 10.1103/PhysRevB.88.104509} {\bibfield
  {journal} {\bibinfo  {journal} {Phys. Rev. B}\ }\textbf {\bibinfo {volume}
  {88}},\ \bibinfo {pages} {104509} (\bibinfo {year} {2013})}\BibitemShut
  {NoStop}%
\bibitem [{SM()}]{SM}%
  \BibitemOpen
  \href@noop {} {}\bibinfo {note} {See Supplemental Material for the analytical
  calculation of the tower of states in the limit $U_0=0,U_2/t\rightarrow 0$,
  and for additional information on the numerical data (excitation spectrum of
  larger systems, finite size analysis of the ferroquadrupolar structure
  factor, of the level crossings, and of the spin gap).}\BibitemShut {Stop}%
\bibitem [{\citenamefont {Anderson}(1952)}]{anderson1952}%
  \BibitemOpen
  \bibfield  {author} {\bibinfo {author} {\bibfnamefont {P.~W.}\ \bibnamefont
  {Anderson}},\ }\href {\doibase 10.1103/PhysRev.86.694} {\bibfield  {journal}
  {\bibinfo  {journal} {Phys. Rev.}\ }\textbf {\bibinfo {volume} {86}},\
  \bibinfo {pages} {694} (\bibinfo {year} {1952})}\BibitemShut {NoStop}%
\bibitem [{\citenamefont {Bernu}\ \emph {et~al.}(1992)\citenamefont {Bernu},
  \citenamefont {Lhuillier},\ and\ \citenamefont {Pierre}}]{bernu1992}%
  \BibitemOpen
  \bibfield  {author} {\bibinfo {author} {\bibfnamefont {B.}~\bibnamefont
  {Bernu}}, \bibinfo {author} {\bibfnamefont {C.}~\bibnamefont {Lhuillier}}, \
  and\ \bibinfo {author} {\bibfnamefont {L.}~\bibnamefont {Pierre}},\ }\href
  {\doibase 10.1103/PhysRevLett.69.2590} {\bibfield  {journal} {\bibinfo
  {journal} {Phys. Rev. Lett.}\ }\textbf {\bibinfo {volume} {69}},\ \bibinfo
  {pages} {2590} (\bibinfo {year} {1992})}\BibitemShut {NoStop}%
\bibitem [{\citenamefont {Bernu}\ \emph {et~al.}(1994)\citenamefont {Bernu},
  \citenamefont {Lecheminant}, \citenamefont {Lhuillier},\ and\ \citenamefont
  {Pierre}}]{bernu1994}%
  \BibitemOpen
  \bibfield  {author} {\bibinfo {author} {\bibfnamefont {B.}~\bibnamefont
  {Bernu}}, \bibinfo {author} {\bibfnamefont {P.}~\bibnamefont {Lecheminant}},
  \bibinfo {author} {\bibfnamefont {C.}~\bibnamefont {Lhuillier}}, \ and\
  \bibinfo {author} {\bibfnamefont {L.}~\bibnamefont {Pierre}},\ }\href
  {\doibase 10.1103/PhysRevB.50.10048} {\bibfield  {journal} {\bibinfo
  {journal} {Phys. Rev. B}\ }\textbf {\bibinfo {volume} {50}},\ \bibinfo
  {pages} {10048} (\bibinfo {year} {1994})}\BibitemShut {NoStop}%
\bibitem [{\citenamefont {Lhuillier}()}]{lhuillier2005}%
  \BibitemOpen
  \bibfield  {author} {\bibinfo {author} {\bibfnamefont {C.}~\bibnamefont
  {Lhuillier}},\ }\href@noop {} {\bibinfo  {journal} {arXiv:cond-mat/0502464}\
  }\BibitemShut {NoStop}%
\bibitem [{\citenamefont {Misguich}\ and\ \citenamefont
  {Sindzingre}(2007)}]{misguich2007}%
  \BibitemOpen
\bibfield  {journal} {  }\bibfield  {author} {\bibinfo {author} {\bibfnamefont
  {G.}~\bibnamefont {Misguich}}\ and\ \bibinfo {author} {\bibfnamefont
  {P.}~\bibnamefont {Sindzingre}},\ }\href
  {http://stacks.iop.org/0953-8984/19/i=14/a=145202} {\bibfield  {journal}
  {\bibinfo  {journal} {Journal of Physics: Condensed Matter}\ }\textbf
  {\bibinfo {volume} {19}},\ \bibinfo {pages} {145202} (\bibinfo {year}
  {2007})}\BibitemShut {NoStop}%
\bibitem [{\citenamefont {Ho}\ and\ \citenamefont {Yip}(2000)}]{ho2000}%
  \BibitemOpen
  \bibfield  {author} {\bibinfo {author} {\bibfnamefont {T.-L.}\ \bibnamefont
  {Ho}}\ and\ \bibinfo {author} {\bibfnamefont {S.~K.}\ \bibnamefont {Yip}},\
  }\href {\doibase 10.1103/PhysRevLett.84.4031} {\bibfield  {journal} {\bibinfo
   {journal} {Phys. Rev. Lett.}\ }\textbf {\bibinfo {volume} {84}},\ \bibinfo
  {pages} {4031} (\bibinfo {year} {2000})}\BibitemShut {NoStop}%
\bibitem [{\citenamefont {Koashi}\ and\ \citenamefont
  {Ueda}(2000)}]{koashi2000}%
  \BibitemOpen
  \bibfield  {author} {\bibinfo {author} {\bibfnamefont {M.}~\bibnamefont
  {Koashi}}\ and\ \bibinfo {author} {\bibfnamefont {M.}~\bibnamefont {Ueda}},\
  }\href {\doibase 10.1103/PhysRevLett.84.1066} {\bibfield  {journal} {\bibinfo
   {journal} {Phys. Rev. Lett.}\ }\textbf {\bibinfo {volume} {84}},\ \bibinfo
  {pages} {1066} (\bibinfo {year} {2000})}\BibitemShut {NoStop}%
\bibitem [{\citenamefont {Mueller}\ \emph {et~al.}(2006)\citenamefont
  {Mueller}, \citenamefont {Ho}, \citenamefont {Ueda},\ and\ \citenamefont
  {Baym}}]{mueller2006}%
  \BibitemOpen
  \bibfield  {author} {\bibinfo {author} {\bibfnamefont {E.~J.}\ \bibnamefont
  {Mueller}}, \bibinfo {author} {\bibfnamefont {T.-L.}\ \bibnamefont {Ho}},
  \bibinfo {author} {\bibfnamefont {M.}~\bibnamefont {Ueda}}, \ and\ \bibinfo
  {author} {\bibfnamefont {G.}~\bibnamefont {Baym}},\ }\href {\doibase
  10.1103/PhysRevA.74.033612} {\bibfield  {journal} {\bibinfo  {journal} {Phys.
  Rev. A}\ }\textbf {\bibinfo {volume} {74}},\ \bibinfo {pages} {033612}
  (\bibinfo {year} {2006})}\BibitemShut {NoStop}%
\bibitem [{\citenamefont {Tasaki}(2013)}]{Tasaki}%
  \BibitemOpen
  \bibfield  {author} {\bibinfo {author} {\bibfnamefont {H.}~\bibnamefont
  {Tasaki}},\ }\href {\doibase 10.1103/PhysRevLett.110.230402} {\bibfield
  {journal} {\bibinfo  {journal} {Phys. Rev. Lett.}\ }\textbf {\bibinfo
  {volume} {110}},\ \bibinfo {pages} {230402} (\bibinfo {year}
  {2013})}\BibitemShut {NoStop}%
\bibitem [{\citenamefont {Sarlo}\ \emph {et~al.}(2013)\citenamefont {Sarlo},
  \citenamefont {Shao}, \citenamefont {Corre}, \citenamefont {Zibold},
  \citenamefont {Jacob}, \citenamefont {Dalibard},\ and\ \citenamefont
  {Gerbier}}]{gerbier2013}%
  \BibitemOpen
  \bibfield  {author} {\bibinfo {author} {\bibfnamefont {L.~D.}\ \bibnamefont
  {Sarlo}}, \bibinfo {author} {\bibfnamefont {L.}~\bibnamefont {Shao}},
  \bibinfo {author} {\bibfnamefont {V.}~\bibnamefont {Corre}}, \bibinfo
  {author} {\bibfnamefont {T.}~\bibnamefont {Zibold}}, \bibinfo {author}
  {\bibfnamefont {D.}~\bibnamefont {Jacob}}, \bibinfo {author} {\bibfnamefont
  {J.}~\bibnamefont {Dalibard}}, \ and\ \bibinfo {author} {\bibfnamefont
  {F.}~\bibnamefont {Gerbier}},\ }\href
  {http://stacks.iop.org/1367-2630/15/i=11/a=113039} {\bibfield  {journal}
  {\bibinfo  {journal} {New Journal of Physics}\ }\textbf {\bibinfo {volume}
  {15}},\ \bibinfo {pages} {113039} (\bibinfo {year} {2013})}\BibitemShut
  {NoStop}%
\bibitem [{\citenamefont {Kimura}\ \emph {et~al.}(2005)\citenamefont {Kimura},
  \citenamefont {Tsuchiya},\ and\ \citenamefont {Kurihara}}]{kimura2005}%
  \BibitemOpen
  \bibfield  {author} {\bibinfo {author} {\bibfnamefont {T.}~\bibnamefont
  {Kimura}}, \bibinfo {author} {\bibfnamefont {S.}~\bibnamefont {Tsuchiya}}, \
  and\ \bibinfo {author} {\bibfnamefont {S.}~\bibnamefont {Kurihara}},\ }\href
  {\doibase 10.1103/PhysRevLett.94.110403} {\bibfield  {journal} {\bibinfo
  {journal} {Phys. Rev. Lett.}\ }\textbf {\bibinfo {volume} {94}},\ \bibinfo
  {pages} {110403} (\bibinfo {year} {2005})}\BibitemShut {NoStop}%
\bibitem [{\citenamefont {Pai}\ \emph {et~al.}(2008)\citenamefont {Pai},
  \citenamefont {Sheshadri},\ and\ \citenamefont {Pandit}}]{pai2008}%
  \BibitemOpen
  \bibfield  {author} {\bibinfo {author} {\bibfnamefont {R.~V.}\ \bibnamefont
  {Pai}}, \bibinfo {author} {\bibfnamefont {K.}~\bibnamefont {Sheshadri}}, \
  and\ \bibinfo {author} {\bibfnamefont {R.}~\bibnamefont {Pandit}},\ }\href
  {\doibase 10.1103/PhysRevB.77.014503} {\bibfield  {journal} {\bibinfo
  {journal} {Phys. Rev. B}\ }\textbf {\bibinfo {volume} {77}},\ \bibinfo
  {pages} {014503} (\bibinfo {year} {2008})}\BibitemShut {NoStop}%
\bibitem [{\citenamefont {Toga}\ \emph {et~al.}(2012)\citenamefont {Toga},
  \citenamefont {Tsuchiura}, \citenamefont {Yamashita}, \citenamefont {Inaba},\
  and\ \citenamefont {Yokoyama}}]{toga2012}%
  \BibitemOpen
  \bibfield  {author} {\bibinfo {author} {\bibfnamefont {Y.}~\bibnamefont
  {Toga}}, \bibinfo {author} {\bibfnamefont {H.}~\bibnamefont {Tsuchiura}},
  \bibinfo {author} {\bibfnamefont {M.}~\bibnamefont {Yamashita}}, \bibinfo
  {author} {\bibfnamefont {K.}~\bibnamefont {Inaba}}, \ and\ \bibinfo {author}
  {\bibfnamefont {H.}~\bibnamefont {Yokoyama}},\ }\href {\doibase
  10.1143/JPSJ.81.063001} {\bibfield  {journal} {\bibinfo  {journal} {Journal
  of the Physical Society of Japan}\ }\textbf {\bibinfo {volume} {81}},\
  \bibinfo {pages} {063001} (\bibinfo {year} {2012})}\BibitemShut {NoStop}%
\bibitem [{\citenamefont {Batrouni}\ \emph {et~al.}(2009)\citenamefont
  {Batrouni}, \citenamefont {Rousseau},\ and\ \citenamefont
  {Scalettar}}]{batrouni2009}%
  \BibitemOpen
  \bibfield  {author} {\bibinfo {author} {\bibfnamefont {G.~G.}\ \bibnamefont
  {Batrouni}}, \bibinfo {author} {\bibfnamefont {V.~G.}\ \bibnamefont
  {Rousseau}}, \ and\ \bibinfo {author} {\bibfnamefont {R.~T.}\ \bibnamefont
  {Scalettar}},\ }\href {\doibase 10.1103/PhysRevLett.102.140402} {\bibfield
  {journal} {\bibinfo  {journal} {Phys. Rev. Lett.}\ }\textbf {\bibinfo
  {volume} {102}},\ \bibinfo {pages} {140402} (\bibinfo {year}
  {2009})}\BibitemShut {NoStop}%
\bibitem [{\citenamefont {Katsura}\ and\ \citenamefont
  {Tasaki}(2013)}]{katsura2013}%
  \BibitemOpen
  \bibfield  {author} {\bibinfo {author} {\bibfnamefont {H.}~\bibnamefont
  {Katsura}}\ and\ \bibinfo {author} {\bibfnamefont {H.}~\bibnamefont
  {Tasaki}},\ }\href {\doibase 10.1103/PhysRevLett.110.130405} {\bibfield
  {journal} {\bibinfo  {journal} {Phys. Rev. Lett.}\ }\textbf {\bibinfo
  {volume} {110}},\ \bibinfo {pages} {130405} (\bibinfo {year}
  {2013})}\BibitemShut {NoStop}%
\bibitem [{\citenamefont {Shannon}\ \emph {et~al.}(2006)\citenamefont
  {Shannon}, \citenamefont {Momoi},\ and\ \citenamefont
  {Sindzingre}}]{shannon2006}%
  \BibitemOpen
  \bibfield  {author} {\bibinfo {author} {\bibfnamefont {N.}~\bibnamefont
  {Shannon}}, \bibinfo {author} {\bibfnamefont {T.}~\bibnamefont {Momoi}}, \
  and\ \bibinfo {author} {\bibfnamefont {P.}~\bibnamefont {Sindzingre}},\
  }\href {\doibase 10.1103/PhysRevLett.96.027213} {\bibfield  {journal}
  {\bibinfo  {journal} {Phys. Rev. Lett.}\ }\textbf {\bibinfo {volume} {96}},\
  \bibinfo {pages} {027213} (\bibinfo {year} {2006})}\BibitemShut {NoStop}%
\bibitem [{\citenamefont {Penc}\ and\ \citenamefont
  {L\"auchli}(2011)}]{penc2011}%
  \BibitemOpen
  \bibfield  {author} {\bibinfo {author} {\bibfnamefont {K.}~\bibnamefont
  {Penc}}\ and\ \bibinfo {author} {\bibfnamefont {A.}~\bibnamefont
  {L\"auchli}},\ }\href@noop {} {\bibfield  {journal} {\bibinfo  {journal} {in
  Introduction to Frustrated Magnetism, edited by C. Lacroix, P. Mendels, and
  F. Mila (Springer-Verlag Berlin Heidelberg), Chap. 13, pp. 331–362}\ }
  (\bibinfo {year} {2011})}\BibitemShut {NoStop}%
\bibitem [{\citenamefont {Rousseau}(2008)}]{rousseau2008}%
  \BibitemOpen
  \bibfield  {author} {\bibinfo {author} {\bibfnamefont {V.~G.}\ \bibnamefont
  {Rousseau}},\ }\href {\doibase 10.1103/PhysRevE.78.056707} {\bibfield
  {journal} {\bibinfo  {journal} {Phys. Rev. E}\ }\textbf {\bibinfo {volume}
  {78}},\ \bibinfo {pages} {056707} (\bibinfo {year} {2008})}\BibitemShut
  {NoStop}%
\bibitem [{\citenamefont {Bergkvist}\ \emph {et~al.}(2006)\citenamefont
  {Bergkvist}, \citenamefont {McCulloch},\ and\ \citenamefont
  {Rosengren}}]{bergkvist2006}%
  \BibitemOpen
  \bibfield  {author} {\bibinfo {author} {\bibfnamefont {S.}~\bibnamefont
  {Bergkvist}}, \bibinfo {author} {\bibfnamefont {I.~P.}\ \bibnamefont
  {McCulloch}}, \ and\ \bibinfo {author} {\bibfnamefont {A.}~\bibnamefont
  {Rosengren}},\ }\href {\doibase 10.1103/PhysRevA.74.053419} {\bibfield
  {journal} {\bibinfo  {journal} {Phys. Rev. A}\ }\textbf {\bibinfo {volume}
  {74}},\ \bibinfo {pages} {053419} (\bibinfo {year} {2006})}\BibitemShut
  {NoStop}%
\bibitem [{\citenamefont {Rodr\'iguez}\ \emph {et~al.}(2011)\citenamefont
  {Rodr\'iguez}, \citenamefont {Arg\"uelles}, \citenamefont {Kolezhuk},
  \citenamefont {Santos},\ and\ \citenamefont {Vekua}}]{rodriguez2011}%
  \BibitemOpen
  \bibfield  {author} {\bibinfo {author} {\bibfnamefont {K.}~\bibnamefont
  {Rodr\'iguez}}, \bibinfo {author} {\bibfnamefont {A.}~\bibnamefont
  {Arg\"uelles}}, \bibinfo {author} {\bibfnamefont {A.~K.}\ \bibnamefont
  {Kolezhuk}}, \bibinfo {author} {\bibfnamefont {L.}~\bibnamefont {Santos}}, \
  and\ \bibinfo {author} {\bibfnamefont {T.}~\bibnamefont {Vekua}},\ }\href
  {\doibase 10.1103/PhysRevLett.106.105302} {\bibfield  {journal} {\bibinfo
  {journal} {Phys. Rev. Lett.}\ }\textbf {\bibinfo {volume} {106}},\ \bibinfo
  {pages} {105302} (\bibinfo {year} {2011})}\BibitemShut {NoStop}%
\bibitem [{\citenamefont {Lamacraft}(2010)}]{lamacraft2010}%
  \BibitemOpen
  \bibfield  {author} {\bibinfo {author} {\bibfnamefont {A.}~\bibnamefont
  {Lamacraft}},\ }\href {\doibase 10.1103/PhysRevB.81.184526} {\bibfield
  {journal} {\bibinfo  {journal} {Phys. Rev. B}\ }\textbf {\bibinfo {volume}
  {81}},\ \bibinfo {pages} {184526} (\bibinfo {year} {2010})}\BibitemShut
  {NoStop}%
\end{thebibliography}%


\begin{thebibliography}{3}%
\makeatletter
\providecommand \@ifxundefined [1]{%
 \@ifx{#1\undefined}
}%
\providecommand \@ifnum [1]{%
 \ifnum #1\expandafter \@firstoftwo
 \else \expandafter \@secondoftwo
 \fi
}%
\providecommand \@ifx [1]{%
 \ifx #1\expandafter \@firstoftwo
 \else \expandafter \@secondoftwo
 \fi
}%
\providecommand \natexlab [1]{#1}%
\providecommand \enquote  [1]{``#1''}%
\providecommand \bibnamefont  [1]{#1}%
\providecommand \bibfnamefont [1]{#1}%
\providecommand \citenamefont [1]{#1}%
\providecommand \href@noop [0]{\@secondoftwo}%
\providecommand \href [0]{\begingroup \@sanitize@url \@href}%
\providecommand \@href[1]{\@@startlink{#1}\@@href}%
\providecommand \@@href[1]{\endgroup#1\@@endlink}%
\providecommand \@sanitize@url [0]{\catcode `\\12\catcode `\$12\catcode
  `\&12\catcode `\#12\catcode `\^12\catcode `\_12\catcode `\%12\relax}%
\providecommand \@@startlink[1]{}%
\providecommand \@@endlink[0]{}%
\providecommand \url  [0]{\begingroup\@sanitize@url \@url }%
\providecommand \@url [1]{\endgroup\@href {#1}{\urlprefix }}%
\providecommand \urlprefix  [0]{URL }%
\providecommand \Eprint [0]{\href }%
\providecommand \doibase [0]{http://dx.doi.org/}%
\providecommand \selectlanguage [0]{\@gobble}%
\providecommand \bibinfo  [0]{\@secondoftwo}%
\providecommand \bibfield  [0]{\@secondoftwo}%
\providecommand \translation [1]{[#1]}%
\providecommand \BibitemOpen [0]{}%
\providecommand \bibitemStop [0]{}%
\providecommand \bibitemNoStop [0]{.\EOS\space}%
\providecommand \EOS [0]{\spacefactor3000\relax}%
\providecommand \BibitemShut  [1]{\csname bibitem#1\endcsname}%
\let\auto@bib@innerbib\@empty
\bibitem [{\citenamefont {Liang}(1995)}]{Liang199511}%
  \BibitemOpen
  \bibfield  {author} {\bibinfo {author} {\bibfnamefont {S.}~\bibnamefont
  {Liang}},\ }\href {\doibase 10.1016/0010-4655(95)00108-R} {\bibfield
  {journal} {\bibinfo  {journal} {Computer Physics Communications}\ }\textbf
  {\bibinfo {volume} {92}},\ \bibinfo {pages} {11 } (\bibinfo {year}
  {1995})}\BibitemShut {NoStop}%
\bibitem [{\citenamefont {Lin}(1990)}]{PhysRevB.42.6561}%
  \BibitemOpen
  \bibfield  {author} {\bibinfo {author} {\bibfnamefont {H.~Q.}\ \bibnamefont
  {Lin}},\ }\href {\doibase 10.1103/PhysRevB.42.6561} {\bibfield  {journal}
  {\bibinfo  {journal} {Phys. Rev. B}\ }\textbf {\bibinfo {volume} {42}},\
  \bibinfo {pages} {6561} (\bibinfo {year} {1990})}\BibitemShut {NoStop}%
\bibitem [{\citenamefont {de~Forges~de Parny}\ \emph
  {et~al.}(2013)\citenamefont {de~Forges~de Parny}, \citenamefont {H\'ebert},
  \citenamefont {Rousseau},\ and\ \citenamefont
  {Batrouni}}]{PhysRevB.88.104509}%
  \BibitemOpen
  \bibfield  {author} {\bibinfo {author} {\bibfnamefont {L.}~\bibnamefont
  {de~Forges~de Parny}}, \bibinfo {author} {\bibfnamefont {F.}~\bibnamefont
  {H\'ebert}}, \bibinfo {author} {\bibfnamefont {V.~G.}\ \bibnamefont
  {Rousseau}}, \ and\ \bibinfo {author} {\bibfnamefont {G.~G.}\ \bibnamefont
  {Batrouni}},\ }\href {\doibase 10.1103/PhysRevB.88.104509} {\bibfield
  {journal} {\bibinfo  {journal} {Phys. Rev. B}\ }\textbf {\bibinfo {volume}
  {88}},\ \bibinfo {pages} {104509} (\bibinfo {year} {2013})}\BibitemShut
  {NoStop}%
\end{thebibliography}%

\end{document}


\title{
 Supplemental Material: Anderson tower of states and nematic order of spin-1 bosonic atoms on a 2D lattice
}

\author{Laurent de Forges de Parny$^{1,2}$, Hongyu Yang$^1$, and Fr\'ed\'eric Mila$^1$}
\affiliation{$^1$ Institut de th\'eorie des ph\'enom\`enes physiques, \'Ecole Polytechnique F\'ed\'erale de Lausanne (EPFL), CH-1015 Lausanne, Switzerland}
\affiliation{$^2$ Laboratoire de Physique, \'Ecole Normale Sup\'erieure de Lyon, 46 All\'ee d'Italie, 69364 Lyon Cedex 07, France}

\date{\today}

\pacs{ 
 05.30.Jp,     
 03.75.Hh,    
 75.10.Jm,     
 03.75.Mn    
}

\maketitle

\section{Analytical investigation of the tower of states in the limit $U_0=0,U_2/t\rightarrow 0$.}

In this section, we derive the expression given in the main text for the coefficients $c_N(S)$ of the expansion of the polar state and 
the expectation value of $\vec S_i^2$ 
in the low-energy states of the Hamiltonian 
\begin{eqnarray}
{\cal H}&=&-t\sum_{\langle i,j\rangle,\sigma} (a^\dagger_{i,\sigma} a^{\vphantom{\dagger}}_{j,\sigma} +\text {H.c.})
+\frac{U_0}{2}\sum_i n_i (n_i-1) \nonumber \\
&+&\frac{U_2}{2} \sum_i (\vec S_i^2-2n_i)
\label{Hamiltonian}
\end{eqnarray}
in the limit $U_0=0,U_2/t\rightarrow 0$. 

To simplify notations, 
we introduce an integer $n$ related to $S_{\text{tot}}$ by
\begin{equation*}
S_{\text{tot}}=N-n
\end{equation*}
where $N$ is the number of bosons, and we define
\begin{equation*}
\vert N-n \rangle\equiv \vert S_{\text{tot}},m=S_{\text{tot}}\rangle
\end{equation*}
In addition, since the ground state manifold is only built out of states of zero momentum, we drop the zero momentum from
the Fourier transforms and define
\begin{equation*}
a^\dagger_\sigma \equiv a^\dagger_{\vec k=\vec 0,\sigma}=\frac{1}{\sqrt{N_s}} \sum_{i=1}^{N_s} a^\dagger_{i,\sigma}
\end{equation*}
where $N_s$ is the number of sites.

With the help of the singlet creation operator $a^{\dagger 2}_0-2 a^\dagger_{-1} a^\dagger_1$, it is clear that $\vert N-n \rangle$ is of the form:
\begin{equation*}
\vert N-n \rangle \propto a^{\dagger (N-n)}_1 (a^{\dagger 2}_0-2 a^\dagger_{-1} a^\dagger_1)^{n/2} \vert 0 \rangle
\end{equation*}
In the following, we will need the explicit form obtained after expanding $(a^{\dagger 2}_0-2 a^\dagger_{-1} a^\dagger_1)^{n/2}$.
It is given by
\begin{equation*}
\vert N-n \rangle = \frac{1}{\sqrt{\text{Norm}}} \sum_{p=0}^{n/2} (-2)^p \binom{\frac{n}{2}}{p }a^{\dagger p}_{-1} a^{\dagger (n-2p)}_0  a^{\dagger (N-n+p)}_1 \vert 0 \rangle
\end{equation*}
with
\begin{eqnarray*}
\text{Norm}&=&\sum_{p=0}^{n/2} 2^{2p} {\binom{\frac{n}{2}}{p }}^2 p! (n-2p)! (N-n+p)!\\
&=&\frac{n!!\,(N-n)!\,(2N-n+1)!!}{(2N-2n+1)!!}
\label{norm}
\end{eqnarray*}

\subsection{Calculation of $c_N(S)$}

In this section, we show how to expand the polar state
\begin{equation*}
\vert \psi_{0,N,0} \rangle \equiv \frac{1}{\sqrt{N!}}a^{\dagger N}_0 \vert 0 \rangle
\end{equation*}
into the basis of the eigenstates of $\vec S_{\text{tot}}^2$ and $S_{\text{tot}}^z$. 
Since $\langle \psi_{0,N,0} \vert S_{\text{tot}}^z \vert \psi_{0,N,0} \rangle=0$, one can
write
\begin{equation*}
\vert \psi_{0,N,0} \rangle = \sum_{S=0,2,...,N} c_N(S) \vert S,m=0\rangle
\end{equation*}
The coefficients $c_N(S)$ are given by
\begin{equation*}
c_N(S)= \langle \psi_{0,N,0} \vert S,m=0\rangle
\end{equation*}
Now, the state $\vert S,m=0\rangle$ can be obtained from the state $\vert S,m=S\rangle$ by applying 
the $S$-th power of the lowering operator $S^-$:
\begin{equation*}
\vert S,m=0\rangle = \frac{S^{(-S)}}{\prod_{p=1}^S \sqrt{S(S+1)-p(p-1)}} \vert S,m=S \rangle
\end{equation*}
or, using the notations introduced in the previous section $S=N-n$ and $\vert N-n \rangle = \vert S,m=S \rangle$,
\begin{eqnarray}
&&\vert S=N-n,m=0\rangle = \nonumber \\
&&\frac{S^{-(N-n)}}{\prod_{p=1}^{N-n} \sqrt{(N-n)(N-n+1)-p(p-1)}} \vert N-n \rangle \nonumber
\end{eqnarray}
To calculate $c_N(S)$, we only need to know the coefficient of $a^{\dagger N}_0\vert 0 \rangle$ in $\vert S=N-n,m=0\rangle$.
Since 
\begin{equation*}
\vert N-n \rangle = \frac{1}{\sqrt{\text{Norm}}} \sum_{p=0}^{n/2} (-2)^p \binom{\frac{n}{2}}{p }a^{\dagger p}_{-1} a^{\dagger (n-2p)}_0  a^{\dagger (N-n+p)}_1 \vert 0 \rangle
\end{equation*}
it is clear that the only term that will give a contribution proportional to $a^{\dagger N}_0\vert 0 \rangle$ when acting on this
state with $S^{-(N-n)}$ is the $p=0$ one, so that 
\begin{eqnarray*}
\langle \psi_{0,N,0} \vert S^{-(N-n)}\vert N-n \rangle = \\
\langle \psi_{0,N,0} \vert S^{-(N-n)} \frac{1}{\sqrt{\text{Norm}}}  a^{\dagger n}_0  a^{\dagger (N-n)}_1 \vert 0 \rangle
\end{eqnarray*}
Now, in $S^{-(N-n)} \frac{1}{\sqrt{\text{Norm}}}  a^{\dagger n}_0  a^{\dagger (N-n)}_1 \vert 0 \rangle$, the only term of
$S^{-(N-n)}$ that gives a contribution proportional to $a^{\dagger N}_0\vert 0 \rangle$ is $(\sqrt{2})^{N-n} (a^\dagger_0 a^{\vphantom{\dagger}}_1)^{N-n}$, and
\begin{equation*}
(a^\dagger_0 a^{\vphantom{\dagger}}_1)^{N-n}a^{\dagger n}_0  a^{\dagger (N-n)}_1 \vert 0 \rangle=(N-n)!a^{\dagger N}_0\vert 0 \rangle 
\end{equation*}
The scalar product $\langle \psi_{0,N,0} \vert S,m=0\rangle$ can now be simply evaluated, leading to
\begin{equation*}
c_N(S)=\frac{(\sqrt{2})^{N-n}\sqrt{N!}\ (N-n)!}{\sqrt{\text{Norm}}\prod_{p=1}^{N-n} \sqrt{(N-n)(N-n+1)-p(p-1)}}
\end{equation*}
or, in terms of $S$ rather then $N-n$,
\begin{equation*}
c_N(S)=\frac{2^{S/2}\ \sqrt{N!}\ S!}{\sqrt{\text{Norm}}\prod_{p=1}^S \sqrt{S(S+1)-p(p-1)}}
\end{equation*}
with
\begin{equation*}
\text{Norm}=\frac{(N-S)!!\,S!\,(N+S+1)!!}{(2S+1)!!}
\label{norm}
\end{equation*}

Finally, one can calculate the product that enters $C_N(S)$ :
\begin{equation*}
\prod_{p=1}^{S} (S(S+1)-p(p-1))=(2S)!
\end{equation*}
leading, after some simplifications, to the following explicit expression for the coefficients $c_N(S)$ :
\begin{equation*}
c_N(S)=\sqrt{\frac{(2S+1)\,N!}{(N-S)!!\,(N+S+1)!!}}
\end{equation*}

\subsection{Calculation of $\langle \vec S_i^2 \rangle$}

In terms of bosonic operators, the components of the spin operators can be written as
$S_i^\alpha=a^\dagger_{i,\sigma} S^\alpha_{\sigma,\tau} a^{\vphantom{\dagger}}_{i,\tau}$,
where $S^\alpha,\alpha=x,y,z$ are the spin-1 matrices, and with implicit summation over repeated 
spin indices. Then, introducing the Fourier transform
$a^\dagger_{\vec k,\sigma}=(1/\sqrt{N_s}) \sum_{i=1}^{N_s} e^{i\vec k . \vec R_i}a^\dagger_{i,\sigma}$,
one can write
\begin{equation*}
\sum_iS_i^\alpha = \sum_{\vec k} a^\dagger_{\vec k,\sigma} S^\alpha_{\sigma,\tau} a^{\vphantom{\dagger}}_{\vec k, \tau}
\end{equation*}
\begin{equation*}
(\sum_i\vec S_i)^2= \sum_{\vec k,\vec q} S^\alpha_{\sigma,\tau} S^\alpha_{\mu,\nu} 
a^\dagger_{\vec k,\sigma} a^{\vphantom{\dagger}}_{\vec k, \tau} a^\dagger_{\vec q,\mu} a^{\vphantom{\dagger}}_{\vec q, \nu}
\end{equation*}
\begin{equation*}
\sum_i(\vec S_i)^2= \frac{1}{N_s}\sum_{\vec k,\vec k',\vec q} S^\alpha_{\sigma,\tau} S^\alpha_{\mu,\nu} 
a^\dagger_{\vec k,\sigma} a^{\vphantom{\dagger}}_{\vec k', \tau} a^\dagger_{\vec q,\mu} a^{\vphantom{\dagger}}_{\vec k-\vec k'-\vec q, \nu}
\end{equation*}

Now, since $a^{\vphantom{\dagger}}_{\vec k, \sigma}\vert N-n\rangle=0$ except when $\vec k = \vec 0$, we get:
\begin{eqnarray}
\langle (\sum_i\vec S_i )^2 \rangle & = & S^\alpha_{\sigma,\tau} S^\alpha_{\mu,\nu} 
\langle a^\dagger_{\vec 0,\sigma} a^{\vphantom{\dagger}}_{\vec 0, \tau} a^\dagger_{\vec 0,\mu} a^{\vphantom{\dagger}}_{\vec 0, \nu}\rangle \nonumber \\
& = & S^\alpha_{\sigma,\tau} S^\alpha_{\mu,\nu} 
\langle a^\dagger_{\vec 0,\sigma} [\delta_{\tau,\mu}+a^\dagger_{\vec 0,\mu} a^{\vphantom{\dagger}}_{\vec 0, \tau}] 
a^{\vphantom{\dagger}}_{\vec 0, \nu}\rangle \nonumber \\
& = & 2N + S^\alpha_{\sigma,\tau} S^\alpha_{\mu,\nu} \langle a^\dagger_{\vec 0,\sigma} a^\dagger_{\vec 0,\mu} a^{\vphantom{\dagger}}_{\vec 0, \tau}  
a^{\vphantom{\dagger}}_{\vec 0, \nu}\rangle
\label{stotsquare}
\end{eqnarray}
where all expectation values are taken in the state $\vert N-n\rangle$, and where we have used the identity 
$S^\alpha_{\sigma,\tau} S^\alpha_{\mu,\nu} \delta_{\tau,\mu} = (\vec S^2)_{\sigma,\nu} = 2 \delta_{\sigma,\nu}$.
Similarly, 
\begin{eqnarray}
N_s\langle \sum_i(\vec S_i)^2 \rangle & = & \sum_{\vec q} S^\alpha_{\sigma,\tau} S^\alpha_{\mu,\nu} 
\langle a^\dagger_{\vec 0,\sigma} a^{\vphantom{\dagger}}_{\vec q, \tau} a^\dagger_{\vec q,\mu} a^{\vphantom{\dagger}}_{\vec 0, \nu}\rangle \nonumber \\
& = & \sum_{\vec q} S^\alpha_{\sigma,\tau} S^\alpha_{\mu,\nu} 
\langle a^\dagger_{\vec 0,\sigma} [\delta_{\tau,\mu}+a^\dagger_{\vec q,\mu} a^{\vphantom{\dagger}}_{\vec q, \tau}] 
a^{\vphantom{\dagger}}_{\vec 0, \nu}\rangle \nonumber \\
& = & 2N_sN + S^\alpha_{\sigma,\tau} S^\alpha_{\mu,\nu} \langle a^\dagger_{\vec 0,\sigma} a^\dagger_{\vec 0,\mu} a^{\vphantom{\dagger}}_{\vec 0, \tau}  
a^{\vphantom{\dagger}}_{\vec 0, \nu}\rangle
\label{sisquare}
\end{eqnarray}

Comparing Eqs. (\ref{stotsquare}) and (\ref{sisquare}), we get:
\begin{equation*}
N_s\langle \sum_i(\vec S_i)^2 \rangle = \langle (\sum_i\vec S_i )^2 \rangle +2N(N_s-1)
\end{equation*}
which, since $\langle (\vec S_i)^2 \rangle$ is independent of $i$, and since $\langle (\sum_i\vec S_i )^2 \rangle=S_{\text{tot}}(S_{\text{tot}}+1)$, leads
to
\begin{equation*}
\langle \vec S_i^2 \rangle = \frac{2N(N_s-1)}{N_s^2}+\frac{1}{N_s^2}S_{\text{tot}}(S_{\text{tot}}+1)
\end{equation*}

\section{ Exact diagonalization of the spin-1 boson model}

\subsection{Method and clusters}

\begin{figure}[h]
\centering
\includegraphics[width=1\columnwidth]{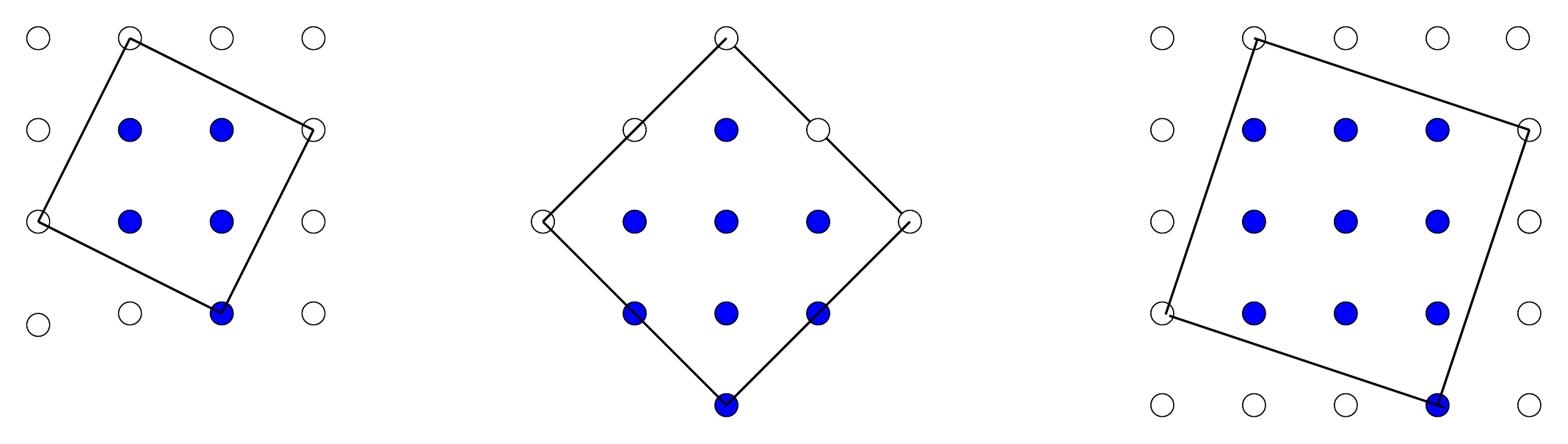}
\caption{(Color online) Sketches of the square clusters with $N_s=5$, $8$ and $10$.}
\label{clusters}
\end{figure}

We have numerically studied the spin-1 boson model on small square clusters by exact diagonalizations. For two bosons per site, we could study clusters with $N_s = 2$, $4$, $5$ and $8$ sites. For one boson per site, an additional cluster with $N_s=10$ has also been studied. The clusters with $N_s=5,8,10$ are depicted in Fig.\ref{clusters}.
The most numerical demanding task was the case with 16 spin-1 bosons on the 8-site cluster.  The $S_z=0$ sector is of dimension 3 567 373 818. We have used the hash function for bosons from Ref.\cite{Liang199511}, and  we have used the ``sublattice coding" trick
\cite{PhysRevB.42.6561} for fast searching.

\subsection{Tower of states}

\begin{figure}[h]
\begin{center}
\includegraphics[width=1\columnwidth]{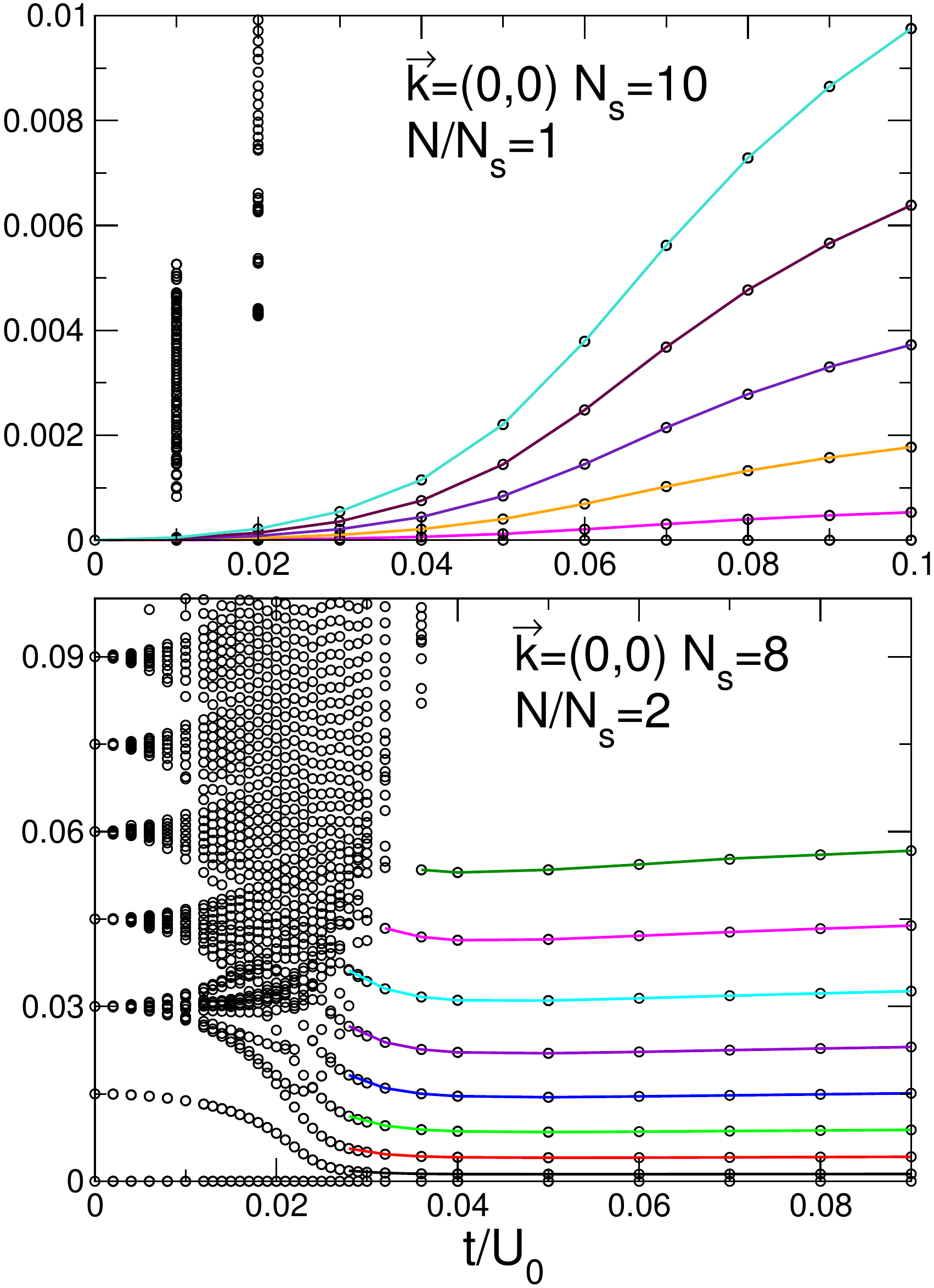}
\caption{(Color online) Low energy spectra in $\vec k=(0,0)$ sector for 10 sites with $N/N_s=1$ and 8 sites with $N/N_s=2$. Both calculations are done for $U_2/U_0=0.005$. }
\label{fig_spectrum}
\end{center}
\end{figure}

The two largest calculations are shown in Fig.\ref{fig_spectrum}. In both cases, one can clearly see the structure of the tower states. 
Here we only show spectrum in the $\vec k=(0,0)$ sector for clarity  since other momenta do not enter the low energy spectrum in the nematic phases. For $N_s=10$ and one boson per site, we can always resolve the five low energy excitations for all nonzero $t/U_0$. For two bosons per site and $N_s=8$, eight excitations can be clearly identified from large to intermediate values of $t/U_0$ until a series of level crossing takes place. 

\begin{figure}[h]
\centering
\includegraphics[width=1\columnwidth]{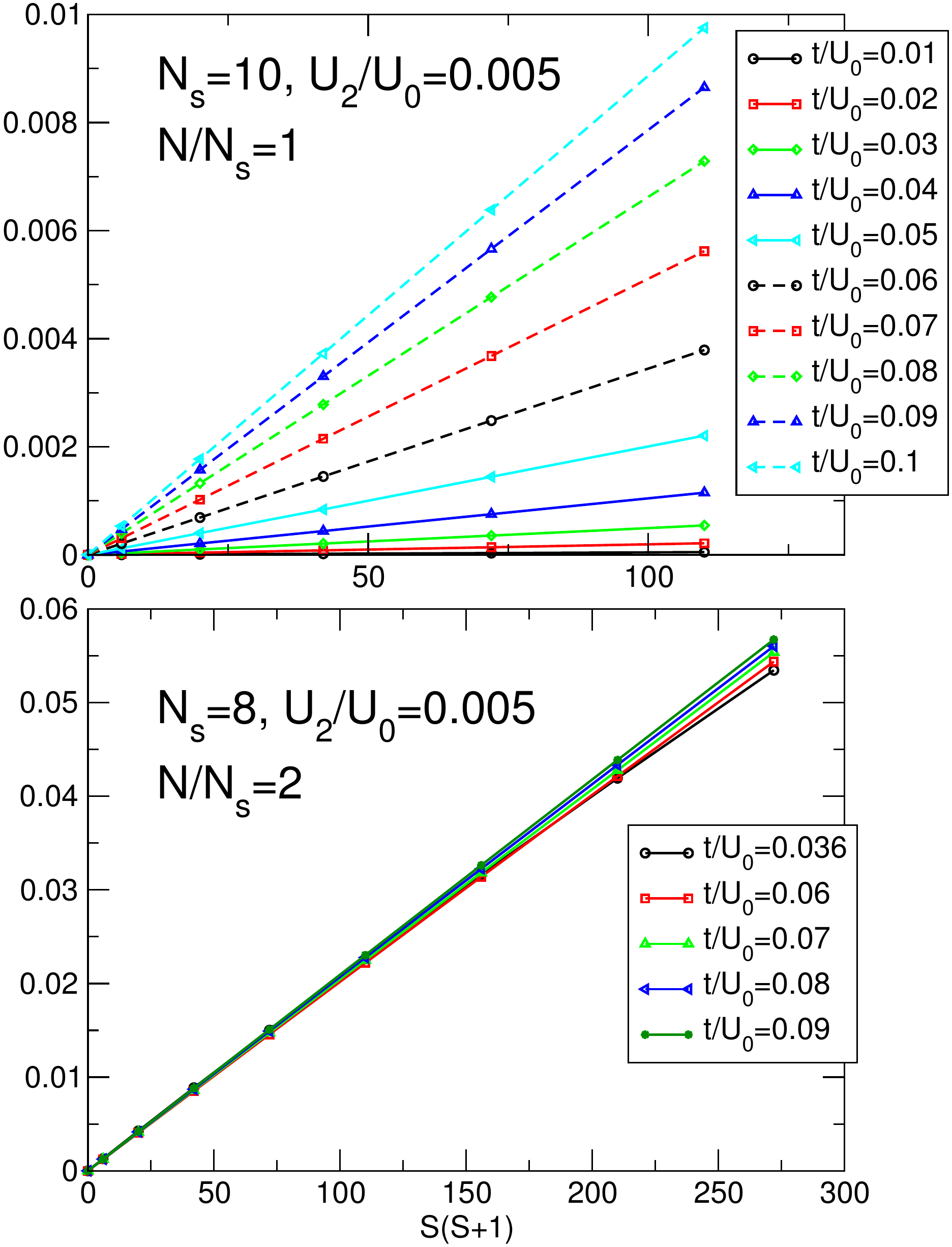}
\caption{(Color online) Plot of the low-lying excitations as a function of $S(S+1)$. The expected linear scaling is manifest.}
\label{tower}
\end{figure}

To check the scaling of these low energy excitations with $S(S+1)$, we have plotted their energy as a function of $S(S+1)$ for several values of $t/U_0$ in Fig.\ref{tower}. As expected, these curves are nearly perfectly linear. Note however that the dependence of the slope on the
ratio $t/U_0$ is quite different in both cases. For one boson per site, the slope goes to zero because all states are continuously connected
to the ground state manifold in the $t/U_0\rightarrow 0$ limit, while for two bosons per site the slope does not change much, a consequence of the fact that the various states are connected to states with different numbers of spin-2 sites in the $t/U_0\rightarrow 0$ limit. 

\subsection{Analysis of the level crossings}

\begin{figure}[h]
\centering
\includegraphics[width=1\columnwidth]{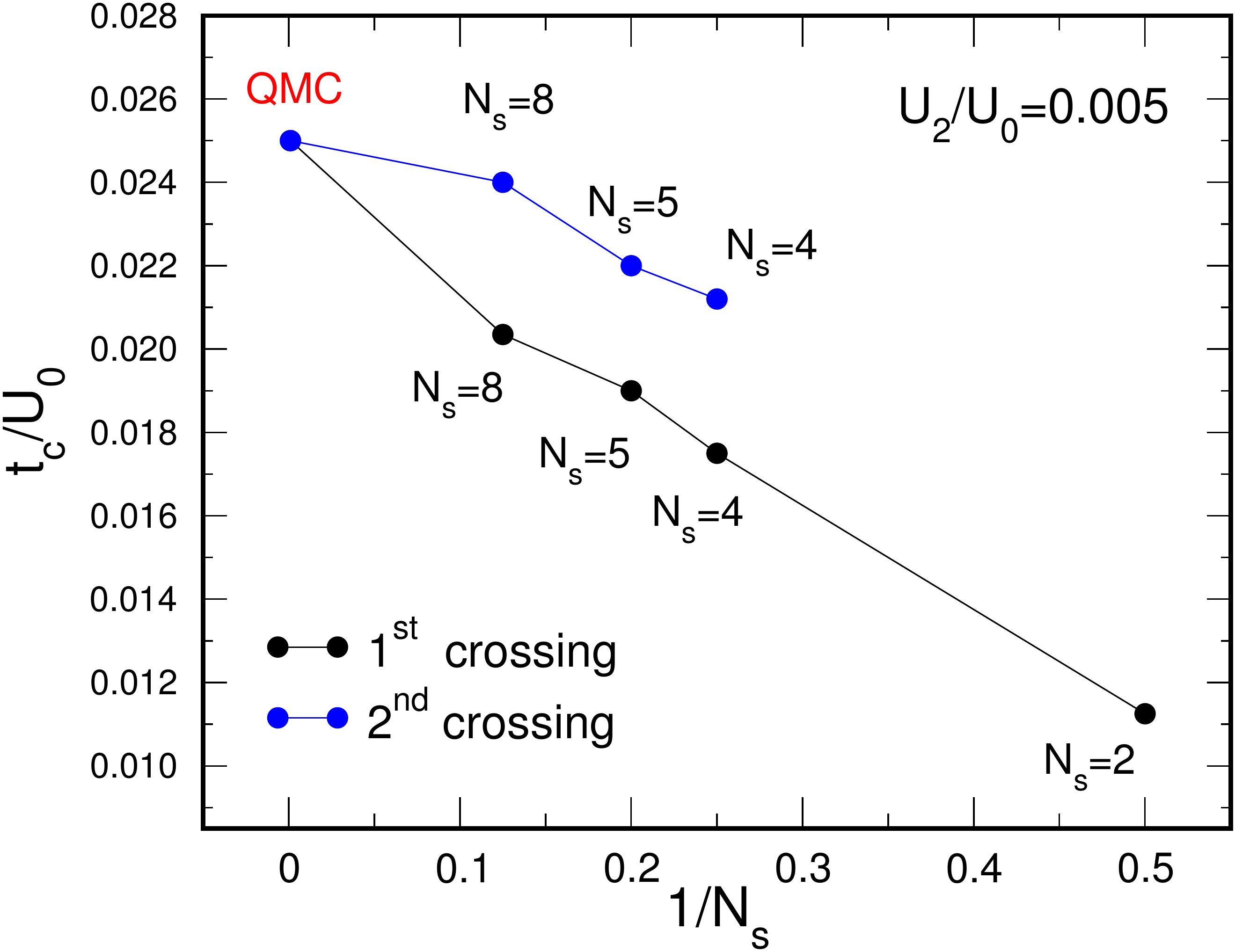}
\caption{(Color online) Finite-size analysis of the first two level crossings for the $N/N_s=2$ case. Their location is consistent with 
the QMC estimate.}
\label{levelcrossing}
\end{figure}

The series of level crossing signals a phase transition and can be used to estimate the critical point. In the singlet phase the
system is expected to have a gap due to the $U_2$ term while the nematic phase is gapless. Thus we have tracked the level crossings from
the singlet phase, and we have determined the location of the first and second level crossings for different sizes as shown in Fig.\ref{levelcrossing}. We do not have any precise analytical prediction for the scaling of these level crossings, but their evolution with
the system size is consistent with the QMC prediction based on the development of a local quadrupole.

\subsection{Ferroquadrupolar structure factor}

\begin{figure}[h]
\centering
\includegraphics[width=1\columnwidth]{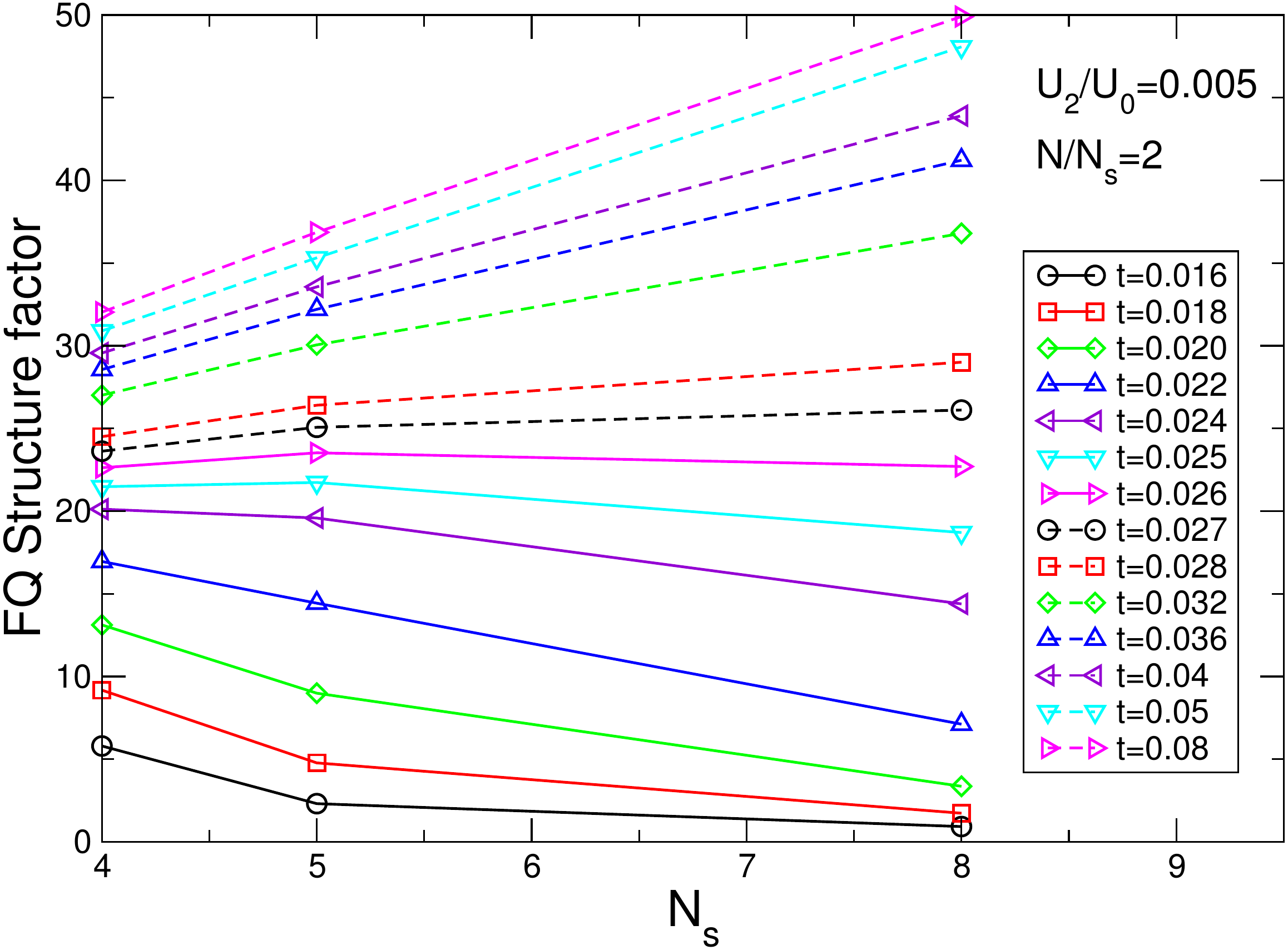}
\caption{(Color online) Scaling of the ferroquadrupolar structure factor with the number of sites $N_s$ for various values of $t/U_0$.}
\label{Structure_Factor}
\end{figure}

To get an alternative indication of the presence of long-range nematic order, 
we have studied the ferroquadrupolar structure factor (see main text). The scaling of the structure factor with the number of sites $N_s$ is
shown in Fig.\ref{Structure_Factor}. It increases linearly with the system size for large enough $t/U_0$, which indicates that there
is indeed nematic long-range order. For $t/U_0\leq 0.026$, by contrast, the structure factor decreases with the number of sites, which
is clearly inconsistent with nematic long-range order. To get a precise estimate of the transition, one should be able to identify the point
below which the structure factor does not diverge any more, which would require to be able to reach much larger system sizes than
available to exact diagonalizations. However, it is plausible that the transition takes place at the point where the structure factor
starts to decrease with the size, i.e. between $t/U_0=0.026$ and $t/U_0=0.027$, again in agreement with the QMC prediction based on the development of a local quadrupole. This change of behavior appears as a level crossing in the figure of the main text.

\subsection{Spin gap}

\begin{figure}[h]
\begin{center}
\includegraphics[width=1\columnwidth]{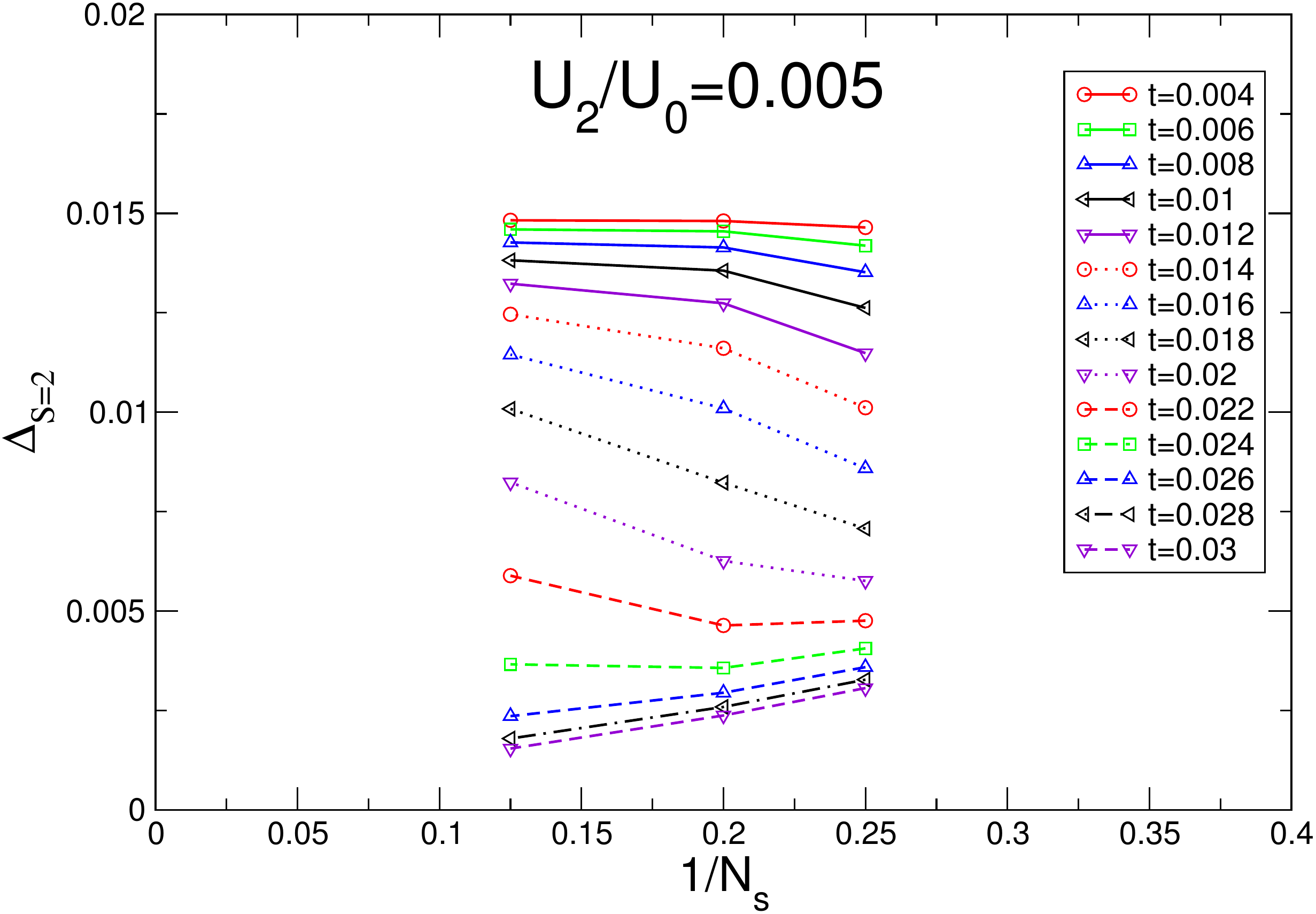}
\caption{(Color online) Scaling of the gap to spin-2 excitations with the number of sites $N_s$ for various values of $t/U_0$.}
\label{fig_gaps}
\end{center}
\end{figure}

Another way to estimate the critical value of $t/U_0$ between the singlet and nematic phases is to calculate spin gaps. Indeed, in the nematic phase, the gap to spin-2 excitations $\Delta_{S=2}$ (the singlet-quintuplet gap) vanishes, while in the singlet phase all gaps remain finite. 
The dependence of $\Delta_{S=2}$ with the size of the clusters is shown in Fig.\ref{fig_gaps} for $U_2/U_0=0.005$. There is a clear
change of behavior for $t/U_0\leq 0.026$, in reasonable agreement with other estimates. However, to be able to determine the precise
value to $t/U_0$ below which the gap vanishes, one should have access to larger sizes.

\subsection{Dependence of the singlet-to-nematic transition on $U_2$}
\begin{figure}[h]
\begin{center}
\includegraphics[width=1\columnwidth]{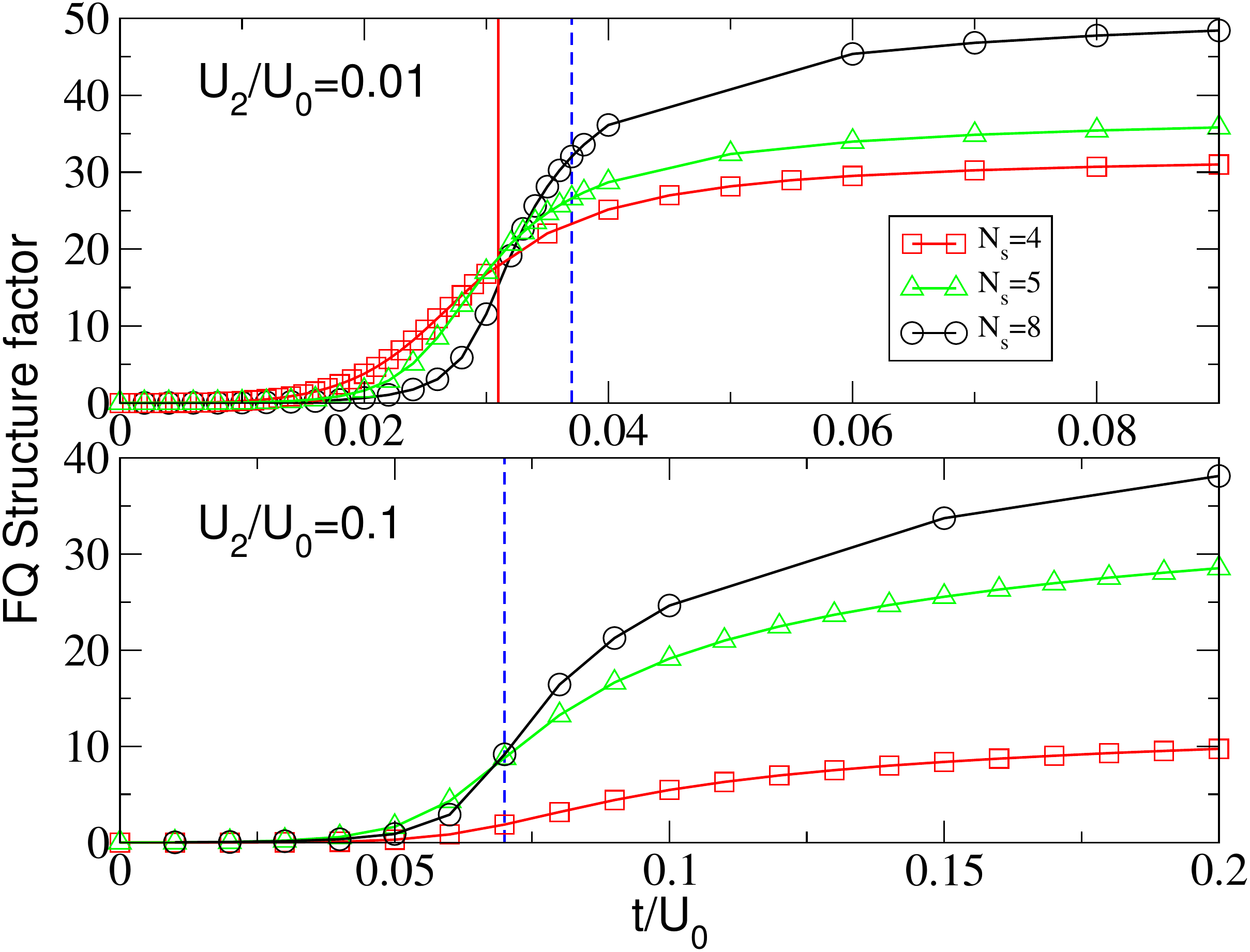}
\caption{(Color online) Ferroquadropolar structure factor for two bosons per site for clusters with $N_s= 4$, $5$, and $8$ sites. 
The vertical dashed (blue) lines indicate the Mott insulator to superfluid transition taken from Ref.\cite{PhysRevB.88.104509} and the vertical straight (red) line indicates the transition between the singlet and nematic phases.}
\label{ferro}
\end{center}
\end{figure}

We have also studied the ferroquadropolar structure factor for other values of $U_2/U_0$ (see Fig.\ref{ferro}). The critical values $t_c/U_0$ derived from this analysis match those estimated by QMC in Ref.\cite{PhysRevB.88.104509} very well. For larger values of $U_2/U_0$ the nematic region inside Mott phase decreases, and for $U_2/U_0=0.1$ it seems that the singlet-to-nematic transition coincides with the Mott-superfluid transition.
\\

\bibliographystyle{apsrev4-1}
\bibliography{suppl}